\documentclass[amsmath,showpacs,nofootinbib,12pt]{revtex4-1}
\usepackage{graphicx}
\usepackage{dcolumn}
\usepackage{bm}
\usepackage{color} 
\usepackage{slashed}
\begin{document}
%\preprint{APS/123-QED}
%%%%%%%%%%%%%%%%%%%%%%%%
\newcommand{\cy}[1]{{\color{cyan}  #1}}
\newcommand{\hs}{\hspace*{0.5cm}}
\newcommand{\vs}{\vspace*{0.5cm}}
\newcommand{\be}{\begin{equation}}
\newcommand{\ee}{\end{equation}}
\newcommand{\bea}{\begin{eqnarray}}
\newcommand{\eea}{\end{eqnarray}}
\newcommand{\ben}{\begin{enumerate}}
\newcommand{\een}{\end{enumerate}}
\newcommand{\bde}{\begin{widetext}}
\newcommand{\ede}{\end{widetext}}
\newcommand{\nn}{\nonumber}
\newcommand{\crn}{\nonumber \\}
\newcommand{\Tr}{\mathrm{Tr}}
\newcommand{\non}{\nonumber}
\newcommand{\noi}{\noindent}
\newcommand{\al}{\alpha}
\newcommand{\la}{\lambda}
\newcommand{\bet}{\beta}
\newcommand{\ga}{\gamma}
\newcommand{\va}{\varphi}
\newcommand{\om}{\omega}
\newcommand{\pa}{\partial}
\newcommand{\+}{\dagger}
\newcommand{\fr}{\frac}
\newcommand{\bc}{\begin{center}}
\newcommand{\ec}{\end{center}}
\newcommand{\Ga}{\Gamma}
\newcommand{\de}{\delta}
\newcommand{\De}{\Delta}
\newcommand{\ep}{\epsilon}
\newcommand{\varep}{\varepsilon}
\newcommand{\ka}{\kappa}
\newcommand{\La}{\Lambda}
\newcommand{\si}{\sigma}
\newcommand{\Si}{\Sigma}
\newcommand{\ta}{\tau}
\newcommand{\up}{\upsilon}
\newcommand{\Up}{\Upsilon}
\newcommand{\ze}{\zeta}
\newcommand{\ps}{\psi}
\newcommand{\Ps}{\Psi}
\newcommand{\ph}{\phi}
\newcommand{\vph}{\varphi}
\newcommand{\Ph}{\Phi}
\newcommand{\Om}{\Omega}
\newcommand{\AdrHEPC}{Phenikaa Institute for Advanced Study and Faculty of Basic Science, Phenikaa University, Yen Nghia, Ha Dong, Hanoi 100000, Vietnam}
%%%%%%%%%%%%%%%%%%%%%%%%

\title{\boldmath Dark charge versus electric charge} 

\author{Duong Van Loi} 
\author{Cao H. Nam} 
\author{Ngo Hai Tan} 
\author{Phung Van Dong} 
\email{Corresponding author.\\ dong.phungvan@phenikaa-uni.edu.vn}
\affiliation{\AdrHEPC} 

\date{\today}
 
\begin{abstract} 
We revisit a theory that proposes a dark charge, $D$, as a dequantization of the electric charge, $Q$. We find that the general arguments of anomaly cancelation and fermion mass generation yield both $D$ and $Q$, nontrivially unified with the weak isospin $T_i$ $(i={1,2,3})$ in a novel gauge symmetry, $SU(3)_C\otimes SU(2)_L\otimes U(1)_Y\otimes U(1)_N$, where $Y$ and $N$ determine $Q$ and $D$ through the $T_3$ operator, i.e. $Q=T_3+Y$ and $D=T_3+N$, respectively. A new observation is that fundamental particles possess a dynamical dark charge which governs both neutrino mass and dark matter, where the neutrino mass is determined via a canonical seesaw, while the dark matter stability is ensured by electric and color charge conservations. We examine the mass spectra of fermions, scalars, and gauge bosons, as well as their interactions, taking into account the kinetic mixing effect of $U(1)_{Y,N}$ gauge fields. The new physics phenomena at colliders are examined. The dark matter relic density and detection are discussed.
\end{abstract}

%\keywords{Neutrino physics, Cosmology of theories beyond the SM, Gauge symmetry}
 
\maketitle

\flushbottom

\section{\label{intro}Motivation}

Neutrino mass \cite{Kajita:2016cak,McDonald:2016ixn} and dark matter \cite{Ade:2015xua,Jungman:1995df,Bertone:2004pz} are the two important questions in science, which cannot be explained within the framework of the standard model. Indeed, the experimental detection of neutrino oscillations has indicated that neutrinos are massive and that flavor lepton numbers are not preserved. In the standard model, neutrinos are massless and flavor lepton numbers are conserved, by contrast. The neutrino oscillations are a clear evidence that the standard model must be extended. Which mechanism produces small neutrino masses and flavor mixing? Further, the standard model content does not contain any candidate for dark matter, which makes up most of the mass of galaxies and galaxy clusters. How is dark matter emerged and stabilized over the cosmological timescales? This work looks for a comprehensive theory which addresses such questions.   

Various theories have been proposed in order to solve both neutrino mass and dark matter, basically given in terms of a seesaw \cite{Minkowski:1977sc,GellMann:1980vs,Yanagida:1979as,Glashow:1979nm,Mohapatra:1979ia,Mohapatra:1980yp,Lazarides:1980nt,Schechter:1980gr,Schechter:1981cv} or/and radiative \cite{Zee:1980ai,Zee:1985id,Babu:1988ki,Krauss:2002px,Ma:2006km} mechanism with the implement of an extra symmetry. Generally, a violation of lepton number \cite{Weinberg:1979sa} would induce appropriate Majorana neutrino masses via the mechanism, whereas the extra symmetry, sometimes interpreted as a residual lepton-number symmetry \cite{Ma:2015xla}, is necessary to make a dark matter candidate stable. Obviously, the lepton symmetry is anomalous, preventing the model's prediction at high energy, while otherwise the extra symmetry, such as a $Z_2$ or matter parity in supersymmetry \cite{Martin:1997ns}, is {\it ad hoc} included, since it is not automatically conserved by the theory. Recent attempts in \cite{Okada:2010wd,Montero:2011jk,Okada:2012sg,Basak:2013cga,Sanchez-Vega:2014rka,Okada:2016tci,Rodejohann:2015lca,Okada:2019sbb,Okada:2020cvq,Dasgupta:2019rmf,Biswas:2019ygr,Gehrlein:2019iwl,Han:2020oet,Choudhury:2020cpm,Mahapatra:2020dgk,Leite:2020wjl,Okada:2020cue,Motz:2020ddk} make use of an anomaly-free abelian gauge symmetry, namely $B-L$ \cite{Davidson:1978pm,Mohapatra:1980qe, Marshak:1979fm}, $L_i-L_j$ \cite{Foot:1990mn,Foot:1994vd,He:1991qd}, or a variant of the weak hypercharge~\cite{Holdom:1985ag,Appelquist:2002mw}. As a result, the model is well defined at high energy and the gauge symmetry breaking leads to appropriate neutrino masses. However, the inclusion of dark matter as well as achieving its stability are still arbitrary. It is therefore desirable to find an underlying principle that determines both neutrino mass generation and dark matter physics.                       

The electric charge ($Q$) of fundamental particles in the nature always comes in discrete amounts, given as integer multiples of a unit, called charge quantization. However, our traditional theories, such as the electrodynamics and the standard model, do not predict this quantization of electric charge. The former theory may imply the charge quantization, if there exists a magnetic monopole as proposed by Dirac long ago \cite{Dirac:1931kp}, but the monopole has not been discovered yet. The latter theory may address the charge quantization if anomaly-free hidden symmetries such as $L_i-L_j$ and $B-L$ are all explicitly violated, since otherwise they make the hypercharge ($Y$), thus the electric charge, free, in such a way that $Y\rightarrow Y+x_{ij}(L_i-L_j)+y(B-L)$ is always allowed, called a dequantization effect \cite{Babu:1989ex,Foot:1990uf}. This work does not solve the question of the charge quantization. By contrast, we argue that the dequantization effect of the electric charge of the standard model might come from the presence of a dark charge, called $D$, which interestingly relates to the neutrino mass and dark matter. The form of the dark charge can be extracted directly as a dequantization of the electric charge, charactered by a $\delta$ parameter, to be achieved in this work. In contrast to the mentioned abelian gauge charges, the dark charge neither commutes nor closes algebraically with the $SU(2)_L$ weak isospin, $T_i$ $(i=1,2,3)$, similarly to the electric charge. Let us note that in \cite{VanDong:2020cjf,VanLoi:2021dzv}, such solutions of the dark charge were applied for further investigations, without derivation and interpretation. 

In deriving the dark charge, the theoretical argument is that the generic hypercharge must be constrained by gauge anomaly cancelation for the model's consistency and gauge-invariant Yukawa Lagrangian for fermion mass generation. As mentioned, the charge quantization in the standard model disappears due to the presence of any anomaly-free hidden symmetry, such as $L_i-L_j$, for $i,j=e,\mu,\tau$, or $B-L$, if one includes three right-handed neutrinos, $\nu_R$'s. We will investigate the latter by imposing $\nu_{R}$'s with $Y(\nu_{R})=\delta$. Solving the conditions of anomaly cancelation and the constraints from Yukawa interactions, we derive a dark charge, $D$, besides the electric charge, to be a natural consequence of the charge dequantization. The condition of algebraic closure between $D$ and $T_i$ demands a novel extension of the standard model gauge symmetry to $SU(3)_C\otimes SU(2)_L\otimes U(1)_Y\otimes U(1)_N$, where $N$ determines the dark charge, $D=T_3+N$, in the same manner in which the hypercharge does so for the electric charge, $Q=T_3+Y$. There is an infinite number of solutions of dark charge symmetries according to distinct values of $\delta$, in which the electric charge is a special case for $\delta=0$. Correspondingly, the gauge completion leads to a model with an infinite number of extra $U(1)_N$ factors. Above, we consider the minimal solution for dark charge, where a unique value for $\delta \neq 0$ defines a nontrivial dark charge, while $\delta=0$ sets the electric charge. In this case, the dark charge $D$, thus $N$, is still arbitrary, but we examine the one according to $\delta=1$, primarily assumed in \cite{VanDong:2020cjf}. 

Conversely, unlike the electric charge, the dark charge anomaly cancelation requires the presence of three right-handed neutrinos, since the usual left-handed neutrinos have a nonzero dark charge. Then the dark charge breaking yields realistic neutrino masses through a canonical seesaw. Additionally, because of the noncommutative nature, the dark charge is broken down to a residual discrete symmetry that divides the standard model fields into several classes, determined by the corresponding values of residual transformations. A new observation of this work is that, although dark fields transform similarly to the usual fields under the residual symmetry, the lightest dark field is stabilized because of the electric and color charge conservations. This is because the lightest dark field is electrically and color neutral, opposite to the charged leptons and quarks. This supplies a dark matter candidate. This feature was also investigated in \cite{VanLoi:2021dzv} when we considered the scenarios of multicomponent dark matter. Since the dark fields interact with the normal fields via $U(1)_N$, the dark dynamics also sets the dark matter observables, besides preventing the dark matter from decay and the role for neutrino mass generation. We will examine the phenomenology of the model in detail. The crucial roles of the dark charge over the electric charge responsible for the new physics are discussed. The new physics effects will be probed through the electroweak precision test, particle colliders, as well as dark matter detections.

The rest of this work is organized as follows. In Sec. \ref{prop}, we reexamine the question of charge quantization when including $\nu_{aR}$, interpreting the dark charge and necessary features of the new model. In Secs. \ref{fer}, \ref{sca}, and \ref{gau}, we investigate the mass spectra of fermion, scalar, and gauge boson, respectively. In Sec. \ref{ints}, we compute necessary interactions of the model. The new physics phenomena and constraints are presented in Secs. \ref{dart}, \ref{dart1}, and \ref{dmnb} corresponding to the electroweak precision test, particle colliders, and dark matter searches, respectively. We conclude this work in Sec. \ref{conl}.

\section{\label{prop} General consideration of the dark charge}

A partial solution of the dark charge was implemented in Ref. \cite{VanDong:2020cjf}. In this section, we derive a generic solution in which the dark charge manifestly arises as a dequantization of the electric charge based upon the general grounds. With this result, we achieve the scheme of single-component dark matter, whereas further implication of the dark charge for multicomponent dark matter is interpreted in Ref. \cite{VanLoi:2021dzv}.   

The electroweak theory is based upon the gauge symmetry, $SU(2)_L\otimes U(1)_Y$. Since the electric charge is additive and conserved, it must be embedded in neutral electroweak charges, such as $Q=\al T_3+\beta Y$. The coefficient $\beta$ can be normalized to 1 because of a scaling symmetry, $g_Y\rightarrow \beta g_Y$ and $Y\rightarrow Y/\beta$, where $g_Y$ is the $U(1)_Y$ coupling, which leaves the theory invariant. The coefficient $\al$ has a dimension of electric charge; and, the $W$ boson has a value of electric charge, $\pm \al$. Since the normalization of $Q$ has not been determined, we use a freedom in assigning the scale of electric charge by fixing the $W$ charge to be unit, i.e. $\al = 1$. Thus, the electric charge in the standard model always takes the form, \be Q=T_3+Y.\ee 

The electric charge is not quantized, because of the form $Q=T_3+Y$. Although $T_3$ is discrete due to the non-abelian nature of the $SU(2)_L$ algebra, the abelian $U(1)_Y$ algebra is trivial, $[Y,Y]=0$, which makes $Y$ arbitrary, in agreement to \cite{Pisano:1996ht,Doff:1998we,deSousaPires:1998jc,deSousaPires:1999ca,VanDong:2005ux}. Notice that $Y$ is often chosen to describe the observed charges while it does not explain them. Further, $Y$ can be constrained by Yukawa Lagrangian and anomaly cancelation, but the standard model might still contain an anomaly-free hidden symmetry, such as $L_i-L_j$, for $i,j=e,\mu,\tau$, or $B-L$, if one includes three right-handed neutrino singlets $\nu_{R}$'s, which subsequently shifts the hypercharge to a generic form, $Y\rightarrow Y+x_{ij}(L_i-L_j)+y(B-L)$, as mentioned. In the following, we consider the latter with $\nu_R$'s and interpret the physics insight.    

Generally, the fermions transform under the electroweak group as 
\bea l_{aL} &=& \left(\begin{array}{c}\nu_{aL}\\ e_{aL}\end{array}\right)\sim (2,Y_{l_a}),\hs \nu_{aR}\sim (1,Y_{\nu_a}),\hs e_{aR}\sim (1,Y_{e_a}),\\ 
q_{aL} &=& \left(\begin{array}{c}u_{aL}\\ d_{aL}\end{array}\right)\sim (2,Y_{q_a}),\hs u_{aR}\sim (1, Y_{u_a}),\hs d_{aR}\sim (1,Y_{d_a}),\eea where we label $a=1,2,3$ to be a generation index. The values in parentheses denote quantum numbers based on $(SU(2)_L,U(1)_Y)$ symmetries, respectively. The right-handed neutrinos $\nu_{aR}$ are introduced, besides the standard model fields, as mentioned. 

The electroweak symmetry breaking and particle mass generation are derived by the Higgs doublet,
\be \phi = \left(\begin{array}{c}\phi_1\\ \phi_2 \end{array}\right)\sim (2,Y_\phi),\ee with a nonzero vacuum expectation value (vev), i.e. $\langle \phi\rangle \neq 0$.
The conservation of electric charge demands that $Q$ must annihilate the weak vacuum, i.e. $Q\langle \phi\rangle=0$, which leads to $Y_\phi =\pm 1/2$. The electric charge of $\phi$ is either $\phi=(\phi^+_1, \phi^0_2)^T$ according to $Y_\phi=1/2$ or $\phi=(\phi^0_1, \phi^-_2)^T$ according to $Y_\phi=-1/2$. Since these solutions yield equivalently physical results, we take $\phi=(\phi^+_1, \phi^0_2)^T \sim (2,1/2)$ with $Y_\phi=1/2$ into account. 

Further, at classical level, the Yukawa Lagrangian, 
\bea \mathcal{L} \supset h^e_{ab} \bar{l}_{aL}\phi e_{bR}+h^\nu_{ab} \bar{l}_{aL}\tilde{\phi} \nu_{bR} + h^d_{ab} \bar{q}_{aL}\phi d_{bR}+h^u_{ab} \bar{q}_{aL}\tilde{\phi} u_{bR}+H.c., \eea must be imposed in order for fermion mass generation and necessary flavor mixings, where we denote $\tilde{\phi}\equiv i\sigma_2 \phi^*$. By the gauge invariance, this Lagrangian gives rise to the hypercharge constraints, such as  
\bea && Y_{q_1}=Y_{q_2}=Y_{q_3}\equiv Y_{q},\hs
Y_{l_1}=Y_{l_2}=Y_{l_3}\equiv Y_{l},\\ 
&& Y_{d_1}=Y_{d_2}=Y_{d_3}\equiv Y_{d},\hs Y_{u_1}=Y_{u_2}=Y_{u_3}\equiv Y_{u},\\
&& Y_{e_1}=Y_{e_2}=Y_{e_3}\equiv Y_{e},\hs Y_{\nu_1}=Y_{\nu_2}=Y_{\nu_3}\equiv Y_{\nu},\\
&& Y_l=Y_\phi+Y_e=-Y_\phi+Y_\nu,\hs Y_q=Y_\phi+Y_d=-Y_\phi+Y_u. \eea
With these conditions at hand, at quantum level, there is only a nontrivial anomaly to be $[SU(2)_L]^2 U(1)_Y$. This anomaly vanishes, if 
\be 3Y_q+Y_l=0.\ee 

With the aid of $Y_\phi=1/2$, the above equations imply \bea &&Y_e=\delta-1,\hs Y_u= 2/3-\delta/3,\hs Y_d=-1/3-\delta/3,\\ 
&&Y_l=-1/2+\delta,\hs Y_q=1/6-\delta/3,\eea which depend on a parameter, $\delta\equiv Y_\nu$. This yields the electric charge of particles, \be Q(\nu)=\delta,\hs Q(e)=\delta-1,\hs Q(u)= 2/3-\delta/3,\hs Q(d)=-1/3-\delta/3,\label{decq} \ee which are not quantized as depending on the $\delta$ parameter.\footnote{If one suppresses $\nu_{aR}$, the anomaly, $[\mathrm{gravity}]^2 U(1)_Y$, is nontrivial and vanishes if $Y_l=-Y_\phi=-1/2$. In this case, the electric charge is quantized, recovered to the observed charges. Alternatively, if one adds a mass term $\nu_{R}\nu_R$ that explicitly violates $B-L$, this leads to the quantization of electric charge again, since $\delta=0$ \cite{Babu:1989tq,Babu:1989ex}.  Obviously, we investigate the universal case in which the neutrinos have right-handed counterparts with $B-L$-conserving Dirac masses as of ordinary fermions.} 

Generally, we have an infinite number of the solutions of hypercharge symmetries corresponding to distinct values of $\delta$, since this parameter is completely arbitrary. Two remarks are given in order, \ben \item True electric charge: $\delta=0$. In this case all the particles get a correct electric charge and hypercharge, as observed, in which $\nu_{aR}$ are a gauge singlet, which can be omitted as in the minimal standard model. The correct electric charge and hypercharge are denoted as $Q\equiv Q|_{\delta=0}$ and $Y\equiv Y|_{\delta=0}$, without confusion.\item Novel dark charge: $\delta\neq 0$. In this case all the particles get an abnormal electric charge (called dark charge) and hypercharge (called hyperdark charge), in which $\nu_{aR}$ are nontrivial under dark charge, which must be included for gravity anomaly cancelation. The dark charge and hyperdark charge are denoted as $D=Q|_{\delta\neq 0}$ and $N=Y|_{\delta\neq 0}$, respectively. This solution differs from the normal one (i.e., the above solution) by a $\delta\neq 0$ for which the dark charge is called a dequantization of the electric charge by a $\delta$ value.\een

Because the two solutions according to $\delta=0$ and $\delta\neq 0$ are linearly independent, i.e. $Y$ and $N$ (thus $Q$ and $D$) are linearly independent, the full gauge symmetry of the theory must take the form, 
\be SU(2)_L\otimes U(1)_Y\otimes U(1)_N,\label{dxep}\ee apart from the QCD group, $SU(3)_C$. Additionally, $Y$ and $N$ determine the electric charge and the dark charge, \be Q=T_3+Y,\hs   D=T_3+N, \label{dtnl110}\ee respectively. It is easily seen that $N$ was identified in the literature as a combination of $x Y+y(B-L)$, for which our model takes $x=1$ and $y=-\delta$, since both $Y,B-L$ are free from anomaly (see, for instance, \cite{Appelquist:2002mw,Oda:2015gna}). However, the dark charge interpretation $D$ as well as this specific combination to be a dequantization of the electric charge have not been presented until this study and a partial solution in \cite{VanDong:2020cjf}. We find the expected relation, \be D=Q-\delta(B-L).\label{adft17}\ee When $\delta\to 0$, then $D\to Q$, meaning that $D$ has properties essentially inherited from $Q$ as a derivation. Whereas, since $\delta$ is finite (i.e., not to infinity), $D$ neither approaches $B-L$ nor regards a commutative nature as of $B-L$ charge. Hence, $D$ is a mirror of $Q$, transformed by $B-L$. Combined with the $T_3$ operator as in (\ref{dtnl110}), the gauge extension (\ref{dxep}) reveals a dark group $U(1)_N$ to be the corresponding mirror of $U(1)_Y$. A crucial result of our approach is that $D$ implies a dark matter stability mechanism and that dark fields may be unified with ordinary fields in weak isospin multiplets, since $D$ is noncommutative.\footnote{Comparable to supersymmetry, superparticle and particle are combined in supermultiplet.} Additionally, although dark fields and normal fields transform nontrivially under the dark charge, the lightest dark field is stable that cannot decay to normal fields, providing a dark matter candidate. This way of dark matter stability differs from that in the most extensions, including $U(1)_{B-L}$, as shown below.    

To make sure, in Appendix \ref{alge}, we investigate another approach that comes to the same conclusion of the gauge symmetry (\ref{dxep}), as desirable. Additionally, all the anomalies vanish, independent of $\delta$, as explicitly verified in Appendix~\ref{anomaly}. In the following, unless otherwise stated, we shall take $\delta=1$ for the case $\delta\neq 0$ into account, which manifestly determines dark matter. Other value of $\delta$ that differs from 1 is viable as studied in \cite{VanLoi:2021dzv}, which would be skipped.  

Each particle (or field) possesses a pair of the characteristic electric and dark charges $(Q,D)$, as collected in Table \ref{tab1}. Notice that the left and right chiral fermions have the same $Q,D$ values, thus their chirality projections have been suppressed. The singlet scalar $\chi$ is necessarily presented to break $U(1)_N$ and generate appropriate right-handed neutrino masses through the coupling $\nu_R\nu_R\chi$, which conserves the dark charge.\footnote{This coupling restricts the electric charge as quantized, because the generated mass $\sim \langle \chi\rangle \nu_R \nu_R$ constrains $Y(\nu_R)=0$. However, the dark charge is always arbitrary, defining $D(\chi)=-2\delta\neq 0$ in the general case, and is broken by $\langle \chi\rangle$. In other words, the dark charge is not only a dequantization version of the electric charge, but also it makes the electric charge quantized.} The last five fields in the table are the gauge fields associated with the gauge symmetry $SU(3)_C \otimes SU(2)_L\otimes U(1)_Y\otimes U(1)_{N}$, where the new field $Z'$ is relevant to $U(1)_N$ extension. The particle representations under this gauge symmetry are listed in Table \ref{tab2}.   
\begin{table}[h]
\bc
\begin{tabular}{c|cccccccccccc}
\hline\hline
Field & $\nu$ & $e$ & $u$ & $d$ & $\phi_1^+$ & $\phi_2^0$ & $\chi$ & gluon & $W^+$ & $A$ & $Z$ & $Z'$ \\
\hline
$Q$ & 0 & $-1$ & $2/3$ & $-1/3$ & 1 & 0 & 0& 0 & 1 & 0& 0& 0 
\\ \hline
$D$ & 1 & 0 & $1/3$ & $-2/3$ & 1 & 0 & $-2$& 0 & 1 & 0&0&0 
\\
\hline
$P_D=(-1)^{kD}$ & $(-1)^k$ & $1$ & $(-1)^{k/3}$ & $(-1)^{-2k/3}$ & $(-1)^k$ & $1$ & 1 & 1 & $(-1)^k$ & 1 & 1 & 1 
\\
\hline\hline
\end{tabular}
\caption[]{\label{tab1} $Q$, $D$ charges and $P_D$ residual symmetry ($k$ integer) of the model particles.}
\ec
\end{table}          
\begin{table}[h]
\bc
\begin{tabular}{c|cccccccc}
\hline\hline
Multiplet & $l_L$ & $q_L$ & $\nu_R$ & $e_R$ & $u_R$ & $d_R$ & $\phi$ & $\chi$ \\
\hline 
$SU(3)_C$ & 1 & 3 & 1 & 1 & 3 & 3 & 1 & 1 \\ \hline  
$SU(2)_L$ & 2 & 2 & 1 & 1& 1& 1& 2 & 1 \\ \hline
$Y$ & $-1/2$ & $1/6$ & 0 & $-1$ & $2/3$ & $-1/3$ & $1/2$ & 0 \\ \hline
$N$ & $1/2$ & $-1/6$ & $1$ & $0$ & $1/3$ & $-2/3$ & $1/2$ & $-2$\\
\hline\hline
\end{tabular}
\caption[]{\label{tab2} $SU(3)_C$, $SU(2)_L$, $Y$, and $N$ quantum numbers of the model multiplets.}
\ec
\end{table} 

The scalars develop the vevs as follows \be \langle \chi\rangle = \fr{1}{\sqrt{2}}\La,\hs \langle \phi \rangle = \fr{1}{\sqrt{2}}\left(\begin{array}{c}
0\\ v\end{array}\right),\ee such that $\La\gg v =246$ GeV to keep a consistency with the standard model.
The scheme of gauge symmetry breaking is \bc \begin{tabular}{c} $SU(3)_C\otimes SU(2)_L\otimes U(1)_Y\otimes U(1)_{N}$ \\
$\downarrow\La$\\
$SU(3)_C\otimes SU(2)_L\otimes U(1)_Y\otimes P_N$\\
$\downarrow v$\\
$SU(3)_C\otimes U(1)_Q\otimes P_D$ \end{tabular}\ec
The $\chi$ vev, $\La$, breaks only $U(1)_N$ down to a residual symmetry, called $P_N$. Next, the weak vacuum, $v$, breaks $SU(2)_L\otimes U(1)_Y\otimes P_N$ down to $U(1)_Q\otimes P_D$, where $Q=T_3+Y$ is as usual, while $P_D$ is a residual symmetry of $D=T_3+N$, shown below. 

To find the explicit form of the residual symmetry, let $X=a_i T_i +b Y + c N$ be the conserved charge after the symmetry breaking. First, it must annihilate the weak vacuum, i.e. $X\langle \phi \rangle =0$, leading to $a_1=a_2=0$ and $a_3=b+c$. Thus, $X=b(T_3+Y)+c(T_3+N)=b Q + c D$. It is clear that $Q$ and $D$ are commuted, i.e. $[Q,D]=0$, and they separately conserve the weak vacuum, $Q\langle \phi \rangle =D\langle \phi \rangle=0$. Hence, the residual symmetry $X$ is abelian, factorized into $U(1)_X=U(1)_Q\otimes U(1)_D$ according to a transformation, $e^{i X}=e^{ib Q}e^{ic D}$. Since $Q$ annihilates the $\chi$ vacuum, $Q\langle \chi \rangle =0$, $U(1)_Q$ is a final residual symmetry, known as the electromagnetic symmetry. Additionally, $U(1)_D$ is a residual symmetry of $SU(2)_L\otimes U(1)_N$ since $D=T_3+N$, which must conserve the $\chi$ vacuum, like $P_N$. Exactly, $cD=X-bQ$ must conserve the $\chi$ vacuum, because both $X,Q$ do. A transformation of $U(1)_D$ is $e^{i c D}$. The vacuum conservation condition demands $e^{i c D} \langle \chi \rangle =\langle \chi \rangle$. It follows that $e^{i c (-2)}=1$, or $c=k \pi$, for $k$ integer. Hence, $U(1)_D$ is reduced to a final residual symmetry, \be P_D=e^{icD} = (-1)^{kD}.\ee It is easily derived, \be P_N=(-1)^{kN},\ee since it is the residual symmetry of $U(1)_N$ that conserves the $\chi$ vacuum, similar to $D$. 

The fundamental difference of the current model from the model with $SU(3)_C\otimes SU(2)_L\otimes U(1)_Y\otimes U(1)_{B-L}$ symmetry structure is the proposal of a noncommutative dark charge $D$, given in (\ref{adft17}). Indeed, the group $U(1)_Y\otimes U(1)_N$ with $N$ to be a combination of $Y$ and $B-L$ is not the same $U(1)_Y\otimes U(1)_{B-L}$ in gauge behavior. The abelian charge $N=Y-\de (B-L)=Q-T_3-\de (B-L)=D-T_3$ is necessarily arisen as a result of the algebraic closure of the dark charge $D$ with $SU(2)_L$ (cf. Appendix \ref{alge}). Additionally, the local symmetry nature of $D$, thus $N$ and their remnants $P_{N,D}$, results from $T_3=D-N$, as $T_3$ is gauged. In contrast, starting from the usual $U(1)_{B-L}$ theory, gauging this group is not required on the theoretical ground, that actually implies a commutative dark parity like the matter parity or a $Z_2$. Even though $U(1)_Y\otimes U(1)_N$ differs from $U(1)_Y\otimes U(1)_{B-L}$ only by the normalization of the $U(1)$ charge, as the group element is just a phase, this phase rotation by $Y$ is not conserved by the weak vacuum, which might lead to distinct physics results for each case. As a matter of fact, a commutative dark parity, analogous to $P_N$, can be realized in the $U(1)_{B-L}$ theory, but a more fundamental dark parity $P_D$ is not given, since $P_D$ only emerges when the weak vacuum possesses a nontrivial hyperdark charge, which necessarily matches $N=Y=1/2$ for $\phi$, as motivated by this work.     

An important remark is that $P_D$ is shifted from $P_N$ by a $SU(2)_L$ transformation, $P_D=P_{T_3} P_N$, determined by the weak breaking, where $P_{T_3}=(-1)^{kT_3}$ contains a weak isospin parity. $P_D$ does not commute with $SU(2)_L$, a consequence of the noncommutative dark charge, i.e., $[D, T_1\pm i T_2]=[T_3, T_1\pm i T_2]=\pm (T_1\pm i T_2)\neq 0$. The difference between $P_N$ and $P_D$ is that $P_N$ commutes with the electroweak symmetry, which transforms every particle in a gauge multiplet identically, whereas $P_D$ transforms component particles that have distinct $T_3$ values differently in a gauge multiplet, thus it separates the components of the weak isospin multiplet. If one extends a known multiplet or introduces a new one, this gives rise to a potential unification of ordinary matter and dark matter with different isospins in the gauge multiplet, in comparison to supersymmetry that does so for particle and superparticle with different spins in a supermultiplet, by contrast. Additionally, unwanted vev directions of a multiplet that have nontrivial $P_D$ are suppressed (e.g., see those in the models extensively discussed in \cite{VanDong:2020cjf}). Hence, $P_D$ is distinct from the usual $U(1)$ extensions, such as $U(1)_{B-L}$, which have only commutative residual symmetry, like the matter parity or $P_N$ in our setup. That said, the approach with noncommutative dark charge would change the current view of neutrino mass and dark matter; for instance, the analysis in \cite {VanDong:2020cjf} yielded that the minimal dark matter, the scotogenic setup, and even the inert Higgs doublet model might be significantly revisited with implement of the dark parity~$P_D$.  

The value of $P_D$ for all fields is collected in Table \ref{tab1}. We deduce that $P_D = 1$ for every field with the minimal $|k| = 6$, except for the identity with $k = 0$. Hence, the residual symmetry $P_D$ is automorphic to
\be Z_6=\lbrace 1,g,g^2,g^3,g^4,g^5\rbrace,\ee 
where $g\equiv (-1)^D$ and $g^6=1$. We factorize $Z_6 \cong Z_2 \otimes Z_3$, where $Z_2 =\lbrace 1, g^3\rbrace$ is the normal subgroup of $Z_6$, while $Z_3 =\lbrace [1],[g^2],[g^4]\rbrace$ is the factor group of $Z_6$ by $Z_2$. Each of $Z_3$ elements contains two elements of $Z_6$, namely $[x]=\{x, g^3 x\}$, hence $[1]=[g^3]=\lbrace 1,g^3\rbrace$, $[g^2]=[g^5]=\lbrace g^2,g^5\rbrace$, and $[g^4]=[g]=\lbrace g,g^4\rbrace$. Since $[g^4]=[g^2]^2=[g^2]^*$ and $[g^2]^3=[1]$, the $Z_3$ group is generated by a generator, \be [g^2]=[\omega^{3D}],\ee where $\omega\equiv e^{i2\pi/3}$ is the cube root of unity. Furthermore, $Z_2$ is generated by a generator, $g^3=(-1)^{3D}$. Since the spin parity $h\equiv (-1)^{2s}$ is always conserved by the Lorentz symmetry, we conveniently multiply $P_D$ with the spin parity group, $P_S=\{1,h\}$, to form $P_D\otimes P_S\cong (Z_2\otimes P_S)\otimes Z_3$. Since $Z_2\otimes P_S$ possesses a normal subgroup, $P=\lbrace 1, p \rbrace$, with    
\be p \equiv g^3\times h=(-1)^{3D+2s},\ee we factorize $P_D\otimes P_S\cong [(Z_2\otimes P_S)/P]\otimes P \otimes Z_3$. Note that $(Z_2\otimes P_S)/P=\{P, \{g^3,h\}\}$ is conserved if $p$, thus $P$, is conserved. We can consider the product,  
\be P\otimes Z_3\subset P_D\otimes P_S, \ee to be the relevant residual symmetry, instead of $P_D$.

The theory conserves both $P$ and $Z_3$ after the symmetry breaking, where $P$ has two irreducible reps, $\underline{1}$ and $\underline{1}'$, according to $p=1$ and $p=-1$, whereas $Z_3$ has three irreducible reps, $\underline{1}$, $\underline{1}'$, and $\underline{1}''$, according to $[g^2]=[1]\rightarrow 1$, $[g^2]=[\omega]\rightarrow \omega$, and $[g^2]=[\om^2]\rightarrow \om^2$, respectively. The reps of $Z_3$ are a homomorphism from $Z_6$, independent of the signs, $g^3=\pm 1$, that identify $Z_6$ elements in a coset. The reps of all fields under $P,Z_{3}$ are given in Table \ref{tabaad1}, where notice that the antiquarks transform as $(\underline{1}')^*=\underline{1}''$ under $Z_3$.\footnote{Reps are always assigned to their group, which should not be confused between reps of $P$ and $Z_3$.}  
\begin{table}[h]
\bc  
\begin{tabular}{c|cccccccccccc}
\hline\hline
Field & $\nu$ & $e$ & $u$ & $d$ & $\phi_1^+$ & $\phi_2^0$ & $\chi$ & gluon & $W^+$ & $A$ & $Z$ & $Z'$ \\
\hline
$p=(-1)^{3D+2s}$ & $1$ & $-1$ & $1$ & $-1$ & $-1$ & 1 & 1 & 1 & $-1$ & 1 & 1 & 1 \\
$[g^2]=[\om^{3D}]$ & 1 & 1 & $\omega$ & $\omega$ & $1$ & 1 & 1 & $1$ & 1 & 1 & 1 & 1\\
$P=\{1,p\}$ & $\underline{1}$ & $\underline{1}'$ & $\underline{1}$ & $\underline{1}'$ & $\underline{1}'$ & $\underline{1}$ & $\underline{1}$ & $\underline{1}$ & $\underline{1}'$ & $\underline{1}$ & $\underline{1}$ & $\underline{1}$ \\
 $Z_3=\{[1],[g^2],[g^4]\}$ & $\underline{1}$ & $\underline{1}$ & $\underline{1}'$ & $\underline{1}'$ & $\underline{1}$ & $\underline{1}$ & $\underline{1}$ & $\underline{1}$ & $\underline{1}$ & $\underline{1}$ & $\underline{1}$ & $\underline{1}$\\
\hline
\end{tabular}
\caption{\label{tabaad1} Field representations under the residual symmetry $P\otimes Z_3$.}
\ec
\end{table}

Under the residual symmetry $P_D$, every dark field introduced should have a dark charge $D$ satisfying $g^6=(-1)^{6D}=1$, hence $3D$ is integer. We derive $D=(3k\pm 1)/3$ or $D=k$, for $k$ integer. The solutions $D=(3k\pm 1)/3$ lead to dark fields that transform nontrivially under $Z_3$ as $[g^2]\rightarrow \om$ or $\om^2$. The lightest $Z_3$ dark field if is color neutral cannot decay to quarks $u,d$ due to the $SU(3)_C$ conservation. (Note that only quarks transform nontrivially under $Z_3$.) Hence, it is stabilized, providing a dark matter candidate. As shown in \cite{VanLoi:2021dzv}, such candidate takes part in multicomponent dark matter scenarios since it is also possibly odd under $P$ and that its stability is only relevant to QCD, which is out of the scope of this work of ``dark charge versus electric charge'' and is omitted. The last solution $D=k$ transforms trivially under $Z_3$ since $[g^2]\rightarrow 1$, but it may be odd under $P$, responsible for dark matter. In the following, we consider only the last solution $D=k$ and note that in this case the theory automatically conserves $Z_3$ due to $SU(3)_C$ symmetry; that is, $Z_3$ acting only on quarks is accidentally preserved by $SU(3)_C$. Omitting the factor group $Z_3$, the residual symmetry is reduced to $P$, and we can redefine, \be P_D = p =(-1)^{3(T_3+N)+2s},\ee called dark parity.\footnote{The dark parity is related to weak isospin, different from those induced by $B-L$, 3-3-1-1, and left-right symmetries \cite{Dong:2013wca,Dong:2014wsa,Alves:2016fqe,Dong:2015yra,Dong:2016sat,Dong:2016gxl,Dong:2017zxo,Huong:2018ytz,Huong:2019vej}.} The dark parity of particles is the $p$ value in Table~\ref{tabaad1}, and we see that the usual fields are divided into two distinct classes, in which $e$, $d$, $\phi_1^+$, and $W^+$ are $P_D$ odd, whereas the rest are $P_D$ even.
      
According to the last solution above, the model can contain a dark field with a dark charge $D=k$, such that $P_D=(-1)^{k+2s}$ is odd. This yields two kinds of candidates: a dark (vectorlike) fermion, labeled $\xi$, for even $k$ and a dark scalar, labeled $\eta$, for odd $k$. They transform under the gauge symmetry $SU(3)_C\otimes SU(2)_L\otimes U(1)_Y\otimes U(1)_{N}$ as \be \xi \sim (1,1,0,2r),\hs \eta\sim (1,1,0,2r-1),\ee for $r$ integer. Here the dark charges are arranged such that the dark fields couple to $\nu_R$ through a Yukawa coupling, $y\bar{\xi}_L \eta\nu_R$, in order to make several dark matter scenarios viable.\footnote{In the case of WIMP dark matter, this coupling is irrelevant and possibly suppressed, hence the dark charge relation might be relaxed. Additionally, when $r=0$, we can introduce only the left chiral component $\xi_L$ (i.e., omitting $\xi_R$), since this field does not contribute to anomaly.} We denote the lightest of $\xi$ and $\eta$ to be $\Psi$. We prove that $\Psi$ can have any mass that does not decay to usual fields. First, $\Psi$, $e$, $d$, $\phi^+_1$, and $W^+$ are all odd. Next, such fields each have an electric or color charge, except for $\Psi$ which is neutral. If $\Psi$ decays, by assumption, the final state has to be electrically and color neutral due to the charge conservations. Since $P_D$ is conserved, the final state must combine an odd number of odd-fields $(e,d,\phi^+_1, W^+)$. Since $W^+$ and $\phi^+_1$ (eaten by $W^+$) decay to $(e^+,\nu)$ and $(d^c,u)$, the final state contains only $(e,d)$ as potential old fields. The decay process looks like, \be \Psi \rightarrow x e^-+\bar{x} e^+ + y d + \bar{y} d^c + z u + \bar{z} u^c+\cdots,\ee where the dots include other fields, if any, which are electrically and color neutral and $P_D$ even. The laws of charge conservations obey, \ben \item $x+\bar{x}+y+\bar{y}=2k+1$ ($P_D$ odd), \item $-x+\bar{x}-y/3+\bar{y}/3+2z/3-2\bar{z}/3=0$ (electrically neutral), \item $y+z-\bar{y}-\bar{z}=3 k'$ (color neutral),\een for $k,k'$ integer. Conditions 2. and 3. give $-x+\bar{x}-y+\bar{y}+2k'=0$, which combined with 1. yields $2(\bar{x}+\bar{y}+k')=2k+1$. This cannot occur, since an even number never equals an odd number. Hence, $P_D$, $U(1)_Q$, and $SU(3)_C$ suppress $\Psi$ decay, if $\Psi$ is heavier than the usual odd fields ($e,d,\phi^+_1,W^+$). $\Psi$ is dark matter and its stability differs from the most extensions; that is, the usual fields transform nontrivially as the dark matter, under the dark parity, but the dark matter stability is preserved by usual electric and color charge conservations.  
              
The total Lagrangian is written as 
\bea \mathcal{L}=\mathcal{L}_{\mathrm{kinetic}}+\mathcal{L}_{\mathrm{Yukawa}}-V. \eea The first part contains kinetic terms and gauge interactions,  
\bea \mathcal{L}_{\mathrm{kinetic}} &=& \sum_F \bar{F}i\ga^\mu D_\mu F+\sum_S (D^\mu S)^\dagger (D_\mu S)\crn
&& -\fr 1 4 G_{m \mu\nu} G_m^{\mu\nu}-\fr 1 4 A_{i \mu\nu} A_i^{\mu\nu}-\fr 1 4 B_{\mu\nu} B^{\mu\nu}-\fr 1 4 C_{\mu\nu} C^{\mu\nu}-\fr{\epsilon}{2} B_{\mu\nu} C^{\mu\nu},\eea 
where $F,S$ run over fermion and scalar multiplets, respectively. The covariant derivative and field strength tensors are defined as   
\bea && D_\mu = \pa_\mu + i g_s t_m G_{m\mu}+ i g T_i A_{i \mu}+ i g_Y Y B_\mu + i g_N N C_\mu,\label{coder}\\
&& G_{m\mu\nu}=\pa_\mu G_{m\nu}-\pa_\nu G_{m\mu}-g_s f_{mpq} G_{p\mu} G_{q\nu},\\
&& A_{i\mu\nu}=\pa_\mu A_{i\nu}-\pa_\nu A_{i\mu}-g \epsilon_{ijk} A_{j\mu} A_{k\nu},\\
&& B_{\mu\nu}=\pa_\mu B_\nu-\pa_\nu B_\mu, \hs C_{\mu\nu}=\pa_\mu C_\nu-\pa_\nu C_\mu, \eea where $(g_s,g,g_Y,g_N)$, $(t_m,T_i,Y,N)$, and $(G_m,A_i,B,C)$ are coupling constants, generators, and gauge bosons according to $(SU(3)_C, SU(2)_L, U(1)_Y, U(1)_{N})$ groups, respectively. And, $f_{mpq}$ and $\epsilon_{ijk}$ are the structure constants of $SU(3)_C$ and $SU(2)_L$, respectively. 

Note that $\epsilon$ is a parameter that determines the kinetic mixing between the two $U(1)$ gauge bosons, satisfying $|\epsilon|<1$, in order for definitely positive kinetic energy. Such kinetic terms can be transformed into the canonical form, i.e. \be -\fr 1 4 B_{\mu\nu} B^{\mu\nu}-\fr 1 4 C_{\mu\nu} C^{\mu\nu}-\fr{\epsilon}{2} B_{\mu\nu} C^{\mu\nu}= -\fr 1 4 \hat{B}_{\mu\nu} \hat{B}^{\mu\nu}-\fr 1 4 \hat{C}_{\mu\nu} \hat{C}^{\mu\nu},\ee by basis changing, 
\begin{eqnarray}
\left(%
\begin{array}{c}
  \hat{B} \\
  \hat{C} \\
\end{array}%
\right)=\left(%
\begin{array}{cc}
  1 & \ep \\
  0 & \sqrt{1-\ep^2} \\
\end{array}%
\right)\left(%
\begin{array}{c}
  B \\
  C \\
\end{array}%
\right).\label{cbad1}
\end{eqnarray}
 
The Yukawa part consists of
\bea \mathcal{L}_{\mathrm{Yukawa}} &=& h^e_{ab} \bar{l}_{aL}\phi e_{bR}+h^\nu_{ab} \bar{l}_{aL}\tilde{\phi} \nu_{bR} + h^d_{ab} \bar{q}_{aL}\phi d_{bR}+h^u_{ab} \bar{q}_{aL}\tilde{\phi} u_{bR}\crn
&& + \fr 1 2 f^\nu_{ab}\bar{\nu}^c_{aR}\chi\nu_{bR} + y_a\bar{\xi}_L\eta\nu_{aR} - m_\xi \bar{\xi}_L \xi_R + H.c., \eea
where $^c$ indicates the charge conjugation, i.e. $\nu^c_R\equiv (\nu_R)^c=C\bar{\nu}^T_R=(\nu^c)_L$, as usual. The scalar potential takes the form 
\bea V &=& \mu^2_1\phi^\dagger \phi + \mu^2_2\eta^* \eta + \mu^2_3 \chi^* \chi + \la_1 (\phi^\dagger \phi)^2 + \la_2(\eta^* \eta)^2 + \la_3 (\chi^* \chi)^2   \crn 
&& + \la_4 (\phi^\dagger \phi)(\eta^* \eta) + \la_5 (\phi^\dagger \phi)(\chi^* \chi) + \la_6 (\eta^* \eta)(\chi^* \chi).\label{potential} \eea 
Note that the couplings $h$'s, $f^\nu$, $y$, and $\la$'s are dimensionless, whereas $m_\xi$ and $\mu$'s have a mass dimension. Especially, when $r=0$, the scalar potential might have extra triple terms, $\mu \chi^*\eta^2 +H.c.$, but they do not affect the present results, hence being neglected.    

\section{\label{fer}Fermion mass}

The spontaneous symmetry breaking will generate fermion masses through the Yukawa Lagrangian. We first consider the charged leptons and quarks, which get
\be [m_e]_{ab}=-h^e_{ab}\fr{v}{\sqrt{2}},\hs [m_u]_{ab}=-h^u_{ab}\fr{v}{\sqrt{2}},\hs [m_d]_{ab}=-h^d_{ab}\fr{v}{\sqrt{2}}.\ee This provides appropriate masses for the particles after diagonalization, similar to the case of the standard model.   

Since the vev of the odd scalar $\eta$ vanishes due to the dark parity conservation, the dark fermion $\xi$ does not mix with right-handed neutrinos $\nu_{aR}$ although they couple via $y_a\bar{\xi}_L\eta \nu_{aR}$. The field $\xi$ is a physical field by itself, with an arbitrary mass $m_\xi$.

The neutrinos $\nu_{a L,R}$ achieve a mass matrix after the two stages of gauge symmetry breaking taking place, such as
\be
\mathcal{L} \supset -\frac{1}{2}\left(%
\begin{array}{cc}
  \bar{\nu}^c_{L} & \bar{\nu}_{R} \\
\end{array}%
\right)
\left(%
\begin{array}{cc}
  0 & m_D \\
  m^T_D & m_M \\
\end{array}%
\right)
\left(% 
\begin{array}{c}
  \nu_{L} \\
  \nu^c_{R} \\
\end{array}
\right)+H.c., \label{adttn12}
\ee
where $[m_D]_{ab}=-h^{\nu*}_{ab} \frac{v}{\sqrt{2}}$ is the Dirac mass matrix that couples $\nu_{aL}$ to $\nu_{bR}$, while $[m_M]_{ab}= -f^{\nu*}_{ab}\frac{\Lambda}{\sqrt{2}}$ is the Majorana mass matrix that couples $\nu_{aR}$ and $\nu_{bR}$ by themselves. 

With the aid of $\La\gg v$, the mass matrix of neutrinos in (\ref{adttn12}) can be diagonalized by a transformation, approximated up to $(v/\La)$ order, to be 
\bea
\left(\begin{array}{c}
\nu_{L} \\
\nu^c_{R} \\
\end{array}
\right)&\simeq&\left(
\begin{array}{cc}
1 & \theta^*  \\
-\theta^T & 1\\
\end{array}\right)\left(
\begin{array}{cc}
U & 0  \\
0 & V^*\\
\end{array}\right)\left(\begin{array}{c}
\nu'_{L} \\
\nu'^c_{R} \\
\end{array}
\right),
\eea where the $\nu_L$-$\nu_R$ mixing element, $\theta= m_D m^{-1}_M\sim v/\La$, is small. The mass eigenvalues kept at $(v/\La)$ order are obtained as
\bea
\mathrm{diag}(m_1,m_2,m_3) &\simeq&-U^T m_Dm^{-1}_Mm^T_DU, \label{nmdd}\\
\mathrm{diag}(M_1,M_2,M_3)&\simeq&V^\dagger m_M V^*,
\eea
where the observed neutrino masses, $m_i \sim v^2/\La$, are appropriately small, while the sterile neutrino masses, $M_i  \sim \La$, are large, at the new physics scale, for $i=1,2,3$ which label the corresponding physical eigenstates, $\nu'_{iL}$ and $\nu'_{iR}$. $U$ is the Pontecorvo-Maki-Nakagawa-Sakata matrix, connecting $\nu_{aL}\simeq U_{ai}\nu'_{iL}$, given that the charged leptons are flavor diagonal, whereas $V$ relates $\nu_{aR}\simeq V_{ai}\nu'_{iR}$. Further, we can take $V=1$ into account, without loss of generality. For convenience, we will omit the prime mark from the physical states, $\nu'_{iL}\rightarrow \nu_{iL}$ and $\nu'_{iR}\rightarrow \nu_{i R}$, without confusion.   

\begin{figure}[h]
\bc
\includegraphics[]{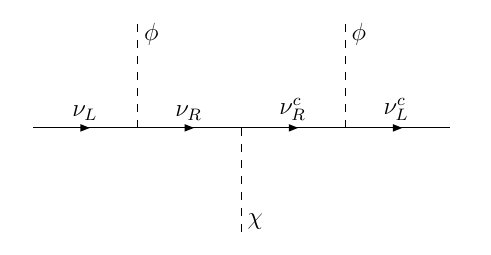}
\caption[]{\label{fig1} Neutrino mass generation seesaw scheme implemented by a dark charge breaking, where $\nu_{L,R}$ carry a unit of dark charge, $D=1$, converted/conserved by the Higgs field $\phi$'s, but then broken by the new Higgs field $\chi$ by two units through a coupling to $\nu_R$'s.}
\ec
\end{figure}

The process of neutrino mass generation is similar to a canonical seesaw, but implemented by a dark charge, instead of the lepton number. This is presented by the flavor diagram in Fig. \ref{fig1}, attached by the external fields $\phi$'s and $\chi$, with the propagations of $\nu_{aR}$. The observed neutrino masses are induced when the dark charge as well as the weak charge are broken by $\langle \chi\rangle $ and $\langle \phi\rangle$, respectively. First, the large Majorana masses $M$ are generated by the interactions of $\nu_{aR}$ with $\chi$ as the middle part in Fig. \ref{fig1}, when the dark charge is broken. In terms of these physical states, the middle part is replaced by a Feynman propagator of $\nu_{iR}$. The observed neutrinos gain small Majorana masses derived by $m_\nu \simeq -m^2_D/M$, when the weak breaking is taking place. It is clear that the full gauge symmetry suppresses all neutrino mass types, but the dark and weak breakings supply desirable neutrino masses, through an improved Higgs mechanism. Last, but not least, this canonical seesaw is naturally realized, since $\nu_{aR}$ appear as fundamental constituents, required by the dark charge symmetry.  Additionally, the Majorana masses of neutrinos emerge from a dark charge breaking, not explicitly relevant to a lepton violation as in the normal sense.

\section{\label{sca}Scalar sector}

Because the electric charge and the dark parity are conserved, only the scalar fields that are electrically neutral and $P_D$ even can develop a vev, such as
$\langle \phi \rangle =\fr{1}{\sqrt{2}}\left(0, v\right)^T$, $\langle \chi\rangle =\fr{1}{\sqrt{2}}\La$, and $\langle \eta\rangle =0$, aforementioned. 

Moreover, necessary conditions for the scalar potential (\ref{potential}) to be bounded from below as well as yielding a desirable vacuum structure are 
\be \la_{1,2,3}>0, \hs \mu^2_{1,3}<0, \hs |\mu_{1}|\ll |\mu_3|, \hs \mu^2_2>0. \ee 

To obtain the potential minimum and physical scalar spectrum, we expand the scalar fields around their vevs as
\bea
&& \phi=\left(\begin{array}{c} \phi^+_1 \\ \frac{1}{\sqrt2}(v+S_1+iA_1) \end{array}\right),\label{expandad} \\ 
&& \chi=\frac{1}{\sqrt2}(\Lambda+S_2+iA_2),\hs \eta=\frac{1}{\sqrt2}(S_3+iA_3),\label{expand}
\eea where note that $\phi^0_2=(v+S_1+i A_1)/\sqrt{2}$.

Substituting (\ref{expandad}) and (\ref{expand}) into (\ref{potential}), the potential minimum conditions are
\be \La^2=\fr{-2\la_5\mu^2_1+4\la_1 \mu^2_3}{\la^2_5-4\la_1\la_3},\hs v^2=\fr{-2\la_5\mu^2_3+4\la_3 \mu^2_1}{\la^2_5-4\la_1\la_3}. \label{addltn01} 
\ee

Using the minimum conditions (\ref{addltn01}), we obtain physical $P_D$-even scalar fields,
\bea
&&\phi=\left(\begin{array}{c} G_{W}^+ \\ \frac{1}{\sqrt2}(v+c_\varphi H + s_\varphi H' +iG_Z) \end{array}\right),\\ 
&&\chi=\frac{1}{\sqrt2}(\Lambda - s_\varphi H + c_\varphi H' +iG_{Z'}), 
\eea
where $G_W\equiv \phi_1$, $G_Z\equiv A_1$, and $G_{Z'}\equiv A_2$ are the massless Goldstone bosons associated with the $W$, $Z$, and $Z'$ gauge bosons, respectively. $H = c_\varphi S_1-s_\varphi S_2$ is identical to the standard model Higgs boson, while $H' = s_\varphi S_1+c_\varphi S_2$ is a new Higgs boson relevant to the dark charge breaking. The $S_1$-$S_2$ mixing angle, $\varphi$, and the $H,H'$ masses are given by
\bea &&t_{2\varphi} =\frac{\la_5 v\Lambda}{\la_3\Lambda^2-\la_1 v^2}\simeq \fr{\la_5}{\la_3}\fr{v}{\La},\\
&&m^2_{H} = \la_1v^2+\la_3\Lambda^2-\sqrt{(\la_1v^2-\la_3\Lambda^2)^2+\la_5^2 v^2\Lambda^2}\simeq \left(2\la_1-\fr{\la^2_5}{2\la_3}\right)v^2,\\
&&m^2_{H'} = \la_1v^2+\la_3\Lambda^2+\sqrt{(\la_1v^2-\la_3\Lambda^2)^2+\la_5^2 v^2\Lambda^2}\simeq 2\la_3\La^2,
\eea which imply that $\varphi$ is small, $m_H$ is at the weak scale, and $m_{H'}$ is at $\La$ scale. 

Last, but not least, the $P_D$-odd fields $S_3,A_3$ do not mix with the $P_D$-even scalars due to the dark parity conservation. $S_3$ and $A_3$ are degenerate in mass, for which they define a physical complex field, say $\eta=(S_3+iA_3)/\sqrt{2}$, with the mass given by 
\be m^2_\eta = \mu_2^2+\fr 1 2\la_4v^2+\fr 1 2\la_6 \Lambda^2.\label{etamm} \ee Depending on the scalar couplings $\la_{4,6}$ and the mass parameter $\mu_2$, the dark scalar $\eta$ can have an arbitrary mass, at $\La$, $v$, or a lower scale.   

\section{\label{gau}Gauge sector}

The gauge bosons acquire masses through their interactions with the scalar fields, when the gauge symmetry breaking happens. The charged gauge boson $W^\pm=(A_1\mp i A_2)/\sqrt{2}$ gets a mass, $m^2_W=g^2v^2/4$, which leads to $v=246$ GeV. 

The mass matrix of the neutral gauge bosons in the canonical basis $(A_{3},\hat{B},\hat{C})^T$, in which the last two are defined in (\ref{cbad1}), is given by
\bea
M^2=L^T_\epsilon\left(%
\begin{array}{ccc}
  \frac{g^2v^2}{4} & -\frac{gg_Yv^2}{4} & -\frac{gg_Nv^2}{4} \\
  -\frac{gg_Yv^2}{4} & \frac{g^2_Yv^2}{4} & \frac{g_Yg_Nv^2}{4} \\
  -\frac{gg_Nv^2}{4} & \frac{g_Yg_Nv^2}{4} & \frac{g^2_Nv^2}{4}+4g^2_N\Lambda^2 \\
\end{array}%
\right)L_\epsilon.\label{addddd}
\eea
Here, note that $L_\epsilon$ is not an orthogonal matrix, relating the canonical basis to the original basis, $(A_3, B, C)^T=L_\epsilon (A_3, \hat{B}, \hat{C})^T$, such that  
\bea
L_\ep=\left(%
\begin{array}{ccc}
  1 & 0 & 0 \\
  0 & 1 & -\fr{\epsilon}{\sqrt{1-\ep^2}} \\
  0 & 0 & \fr{1}{\sqrt{1-\ep^2}}\\
\end{array}%
\right).
\eea Since the usual Higgs field has a hyperdark charge, it produces the mixing mass terms between $(A_3,B)$ and $C$ as given in (\ref{addddd}) due to the electroweak symmetry breaking. Such mixing mass terms vanish in the usual $U(1)_{B-L}$ theory.

It is easily checked that the mass matrix (\ref{addddd}) provides a zero eigenvalue (i.e., the photon mass) with a corresponding eigenstate (i.e., the photon field) to be
\be A=s_W A_3+c_W\hat{B},\ee where the Weinberg angle is defined by $t_W=g_Y/g$.\footnote{Interested reader can refer to \cite{Dong:2015jxa,VanDong:2005pi,VanLoi:2019xud} for diagonalizing a more-general neutral-gauge sector with/without a kinetic mixing term.} The $Z_0$ boson is defined, orthogonal to the photon $A$, such as   
\be Z_0=c_W A_3-s_W\hat{B},\ee which is identical to that of the standard model. Hence, in the new basis $(A, Z_0, \hat{C})^T$, the photon is decoupled, as a physical field, whereas there remains a mixing between $Z_0$ and $\hat{C}$. By diagonalization, the last two yield physical fields, $Z=c_\al Z_0 -s_\al \hat{C}$ and $Z'=s_\al Z_0 +c_\al \hat{C}$, determined through a mixing angle, $\al$, evaluated by 
\bea
t_{2\alpha}\simeq -\frac{\sqrt{1-\ep^2}}{8g^2_N}\sqrt{g^2+g^2_Y}(g_N-\ep g_Y)\frac{v^2}{\Lambda^2}.
\eea 

That said, the mass matrix (\ref{addddd}) is fully diagonalized, \be O^TM^2O=\text{diag}(0,m^2_Z,m^2_{Z'}),\ee by an orthogonal transformation, 
\bea
O=\left(%
\begin{array}{ccc}
  s_W & c_W & 0 \\
  c_W & -s_W & 0 \\
  0 & 0 & 1 \\
\end{array}%
\right)\left(%
\begin{array}{ccc}
  1 & 0 & 0 \\
  0 & c_\alpha & s_\alpha \\
  0 & -s_\alpha & c_\alpha\\
\end{array}%
\right),
\eea which relates the physical states to the canonical states, $(A_{3}, \hat{B}, \hat{C})^T=O (A, Z, Z')^T$. And, the mass eigenvalues are approximated as
\bea
m^2_Z&\simeq &\frac{g^2+g^2_Y}{4}v^2\left[1-\frac{(g_N-\ep g_Y)^2}{16g^2_N}\frac{v^2}{\Lambda^2}\right],\\
m^2_{Z'}&\simeq &\frac{4g^2_N\Lambda^2}{1-\ep^2}\left[1+\frac{(g_N-\ep g_Y)^2}{16g^2_N}\frac{v^2}{\Lambda^2}\right].
\eea 

Note that the $Z$-$Z'$ mixing, i.e. the $\al$ angle, comes from the two sources, the kinetic mixing characterized by $\epsilon$ and the symmetry breaking induced by $v,\La$. Such two contributions cancel out if $\epsilon = g_N/g_Y$. This phenomenon does not exist in the usual $U(1)_{B-L}$ theory. Additionally, the well-measured quantities, such as the $Z$ couplings and the $\rho$-parameter, are modified by the difference $g_N-\ep g_Y$, which occurs even in absence of the kinetic mixing. In the usual $U(1)_{B-L}$ theory, such modifications are proportional to $\ep$ and thus disappear when the kinetic mixing is suppressed, by contrast.  

Note also that the physical states $(A,Z,Z')$ are related to the original states $(A_{3},B,C)$, such as
$(A_{3}, B, C)^T=L_\epsilon O (A, Z, Z')^T$.

\section{\label{ints}Interactions}

We investigate the interactions of electroweak and new gauge bosons with fermions. Let us expand the relevant Lagrangian, 
\be
\sum_F \bar{F}i\ga^\mu D_\mu F = \sum_F \bar{F}i\ga^\mu \partial_\mu F - g_s\sum_F \bar{F}\ga^\mu t_mG_{m\mu} F + \mathcal{L}^{CC} + \mathcal{L}^{NC},
\ee
where 
\bea \mathcal{L}^{CC} &=& - g\sum_{F_L}\bar{F}_L\ga^\mu (T_1A_{1\mu}+T_2A_{2\mu})F_L,\label{LCC}\\
\mathcal{L}^{NC} &=& - g \sum_{F_L}\bar{F}_L\ga^\mu (T_3A_{3\mu}+t_W Y_{F_L}B_\mu+t_NN_{F_L}C_\mu)F_L \crn
&& - g\sum_{F_R}\bar{F}_R\ga^\mu (t_W Y_{F_R}B_\mu+t_NN_{F_R}C_\mu)F_R,\label{LNC}
\eea
where $F_L$ and $F_R$ run over the left-handed and right-handed fermion multiplets of the model, respectively, and we define $t_N=g_N/g$. 

From (\ref{LCC}), we obtain the interactions of fermions with charged gauge bosons,
\be \mathcal{L}^{CC} = -\frac{g}{\sqrt2}(\bar{e}_{L}\ga^\mu U \nu_{L}+\bar{d}_{L}\ga^\mu V_{\mathrm{CKM}}u_{L})W^-_\mu+H.c.,
\ee
where we denote
$\nu \equiv \left(\nu_{1}, \nu_{2}, \nu_{3}\right)^T$, $e\equiv \left(e, \mu, \tau \right)^T$,
$u\equiv \left(u, c, t\right)^T$, and $d\equiv \left(d, s, b \right)^T$ to be mass eigenstates, without confusion. 

Equation (\ref{LNC}) gives rise to the interactions of fermions with neutral gauge bosons,
\bea \mathcal{L}^{NC} &=& -eQ(f)\bar{f}\ga^\mu f A_\mu \crn
&& - \frac{g}{2c_W}\{C^Z_{\nu_L}\bar{\nu}_L\ga^\mu\nu_L + C^Z_{\nu_R}\bar{\nu}_R\ga^\mu\nu_R + \bar{f}\ga^\mu [g_V^Z(f)-g_A^Z(f)\ga_5]f\} Z_\mu \crn
&& - \frac{g}{2c_W}\{C^{Z'}_{\nu_L}\bar{\nu}_L\ga^\mu\nu_L + C^{Z'}_{\nu_R}\bar{\nu}_R\ga^\mu\nu_R+\bar{f}\ga^\mu [g_V^{Z'}(f)-g_A^{Z'}(f)\ga_5]f\} Z'_\mu, 
\eea
where $f$ is summed over every fermion of the model, except for neutrinos, and
\bea C^Z_{\nu_L} &=& c_\al-\frac{c_Wt_N+\ep s_W}{\sqrt{1-\ep^2}}s_\al,\hs C^Z_{\nu_R} = -\frac{2c_Wt_N}{\sqrt{1-\ep^2}}s_\al,\\
C^{Z'}_{\nu_L} &=& s_\al+\frac{c_Wt_N+\ep s_W}{\sqrt{1-\ep^2}}c_\al,\hs C^{Z'}_{\nu_R} = \frac{2c_Wt_N}{\sqrt{1-\ep^2}}c_\al.\label{neuZp}
\eea
The vector and axial-vector couplings of $Z,Z'$ to the remaining fermions are listed in Tables \ref{Z} and \ref{Zp}, respectively. 
\begin{table}[!h]
\bc
\begin{tabular}{c|cc}
\hline\hline
$f$ & $g^Z_V (f)$ & $g^Z_A (f)$\\ 
\hline
$e,\mu,\tau$ & $\frac{1-2c_{2W}}{2}c_\al-\frac{c_Wt_N+3\ep s_W}{2\sqrt{1-\ep^2}}s_\al$ & $-\frac{1}{2}c_\al-\frac{c_Wt_N-\ep s_W}{2\sqrt{1-\ep^2}}s_\al$ \\
$u,c,t$ & $-\frac{1-4c_{2W}}{6}c_\al-\frac{c_Wt_N-5\ep s_W}{6\sqrt{1-\ep^2}}s_\al$ & $\frac{1}{2}c_\al+\frac{c_Wt_N-\ep s_W}{2\sqrt{1-\ep^2}}s_\al$ \\
$d,s,b$ & $-\frac{1+2c_{2W}}{6}c_\al+\frac{5c_W t_N-\ep s_W}{6\sqrt{1-\ep^2}}s_\al$ & $-\frac{1}{2}c_\al-\frac{c_Wt_N-\ep s_W}{2\sqrt{1-\ep^2}}s_\al$ \\
$\xi$ & $-\frac{4r c_W t_N}{\sqrt{1-\ep^2}}s_\al$ & $0$\\
\hline\hline
\end{tabular}
\caption[]{\label{Z}Couplings of $Z$ with fermions ($f\neq \nu$).}
\ec  
\end{table}
\begin{table}[!h]
\bc
\begin{tabular}{c|cc}
\hline\hline
$f$ & $g^{Z'}_V(f)$ & $g^{Z'}_A(f)$\\ 
\hline
$e,\mu,\tau$ & $\frac{1-2c_{2W}}{2}s_\al+\frac{c_Wt_N+3\ep s_W}{2\sqrt{1-\ep^2}}c_\al$ & $-\frac{1}{2}s_\al+\frac{c_Wt_N-\ep s_W}{2\sqrt{1-\ep^2}}c_\al$ \\
$u,c,t$ & $-\frac{1-4c_{2W}}{6}s_\al+\frac{c_Wt_N-5\ep s_W}{6\sqrt{1-\ep^2}}c_\al$ & $\frac{1}{2}s_\al-\frac{c_Wt_N-\ep s_W}{2\sqrt{1-\ep^2}}c_\al$ \\
$d,s,b$ & $-\frac{1+2c_{2W}}{6}s_\al-\frac{5c_Wt_N-\ep s_W}{6\sqrt{1-\ep^2}}c_\al$ & $-\frac{1}{2}s_\al+\frac{c_Wt_N-\ep s_W}{2\sqrt{1-\ep^2}}c_\al$ \\
$\xi$ & $\frac{4r c_Wt_N}{\sqrt{1-\ep^2}}c_\al$ & $0$\\
\hline\hline
\end{tabular}
\caption[]{\label{Zp}Couplings of $Z'$ with fermions ($f\neq \nu$).}  
\ec
\end{table}

Generically, the couplings of $Z$ with fermions deviate from the standard model prediction due to the two sources, the dark charge breaking and the kinetic mixing, as mentioned. However, for the neutrino coupling, we find $C^Z_{\nu_L}\simeq 1+v^2/16\La^2$ at the leading order, which comes only from the dark charge breaking, not from the kinetic mixing. This leading contribution disappears in the usual $U(1)_{B-L}$ theory, which starts from $\mathcal{O}[(v^2/\La^2)\ep]$ by contrast. The contribution causes a deviation from the standard model prediction on invisible $Z$ decay width to neutrinos by an amount, $\Delta \Ga_{\mathrm{inv}}/\Ga_{\mathrm{inv}}\simeq v^2/8\La^2\lesssim 0.005$, where the last number agrees with the electroweak measurement \cite{Tanabashi:2018oca}, which gives $\La\gtrsim 5 v\simeq 1.23$ TeV.

Similar to the standard model $Z$ boson, the couplings of the new $Z'$ boson to ordinary fermions violate parity at a considerable level, since $g^{Z'}_A$ is always nonzero, even at the effective limit $v\ll \La$ and $|\ep|\ll1$. This is due to the fact that the left-handed and right-handed ordinary fermions including neutrinos have different hyperdark charges, as seen from Table \ref{tab2}. In the usual $U(1)_{B-L}$ model, the interactions of the $B-L$ gauge boson to ordinary fermions are almost vectorlike, i.e. conserving parity, for $|\ep|\ll 1$. This is an important feature for discriminating our model and the $U(1)_{B-L}$ model in experiment. Particularly, unlike a vectorlike $B-L$ gauge boson, the $Z'$ boson in our model contributes to atomic parity violation through the effective Lagrangian, \be \mathcal{L}^{Z'}_{\mathrm{eff}}\supset \fr{G_{F}}{\sqrt{2}}(\bar{e}\ga_\mu \ga_5 e)(C'_{1u}\bar{u}\ga^\mu u + C'_{1d} \bar{d}\ga^\mu d),\ee where $G_F/\sqrt{2}=1/2v^2$, $C'_{1u}\simeq v^2/96\La^2$, and $C'_{1d}\simeq -5v^2/96\La^2$. The parity violation for vector-coupled electrons and axially-coupled quarks due to $Z'$ also arises, but suppressed for a heavy atom because of its dependence on spins rather than charges, similar to the $Z$ boson effect; thus, this kind of contribution is neglected. The weak charge deviation from the standard model prediction accounted for an atom consisting of $Z$ protons and $N$ neutrons is obtained as \be \Delta Q_{W}(Z,N) = -2[Z(2C'_{1u}+C'_{1d})+N(C'_{1u}+2C'_{1d})]\simeq \fr{Z+3N}{16}\fr{v^2}{\La^2}\simeq 18\fr{v^2}{\La^2},\ee where the last number is applied for Cesium with $Z=55$ and $N=78$. The current experiment and the standard model prediction for Cs weak charge are supplied in \cite{Tanabashi:2018oca} that makes a bound $\Delta Q_{\mathrm{W}}(\mathrm{Cs})<0.61$, implying $\La>5.43 v\simeq 1.33$ TeV.

The bounds of the $\La$ new physics scale from the invisible $Z$ decay and the Cesium parity violation obviously satisfy the combined constraints studied below, which need not necessarily be included to the final result.

\section{\label{dart}Electroweak precision test} 

\subsection{$\rho$-parameter}
Because the $Z$ boson mixes with the new neutral gauge boson $Z'$ through the kinetic mixing and the symmetry breaking, the new physics contributions to the $\rho$-parameter start from the tree-level, given by
\be \Delta\rho = \frac{m^2_W}{c^2_W m^2_Z}-1\simeq \frac{(t_N-\ep t_W)^2}{16 t^2_N}\frac{v^2}{\Lambda^2}.
\ee
From the global fit, the $\rho$ parameter is bounded by $0.0002 < \Delta\rho < 0.00058$ \cite{Tanabashi:2018oca}, which leads to the following lower bound,
\be \Lambda \gtrsim 2.553 \times\frac{|t_N-\ep t_W|}{t_N}\ \ \text{TeV}.
\ee

In the $U(1)_{B-L}$ model, one has a bound for relevant new physics scale likely $\La\gtrsim 2.553\times |\ep|g_Y/g_{B-L}$ TeV, which is easily evaded for small $|\ep|$, given that $g_Y\sim g_{B-L}$. However, in the current model, even for $|\ep|\ll 1$, the new physics scale is always limited by $\La\gtrsim 2.553$ TeV, by contrast. This bound is quite bigger than those given by the invisible $Z$ decay and the Cs parity violation.

\subsection{Total $Z$ decay width}
We will use the precision measurement of the total $Z$ decay width to impose the constraint on the free parameters of the model. The total $Z$ decay width is measured by the experiment and predicted by the standard model, respectively, by \cite{Tanabashi:2018oca} \be \Gamma^{\text{exp}}_Z=2.4952\pm0.0023\ \mathrm{GeV},\hs \Gamma^{\text{SM}}_Z=2.4942\pm0.0008\ \mathrm{GeV}.\ee

First, we rewrite the Lagrangian describing the $Z$ couplings to the standard model fermions, such as
\bea \mathcal{L}^{NC} &\supset& - \frac{g}{2c_W}\left\{\bar{\nu}_L\ga^\mu(1+\Delta_{\nu_L})\nu_L \right.\crn
&& \left. + \bar{f}\ga^\mu [g_{0V}^Z(f)(1+\Delta_{V,f})-g_{0A}^Z(f)(1+\Delta_{A,f})\ga_5]f\right\} Z_\mu,
\eea
where $g_{0V}^Z(f)=T_3(f)-2Q(f)s^2_W$ and $g_{0A}^Z(f)=T_3(f)$ are the standard model predictions for the vector and axial-vector couplings, respectively. $\Delta_{\nu_L}$, $\Delta_{V,f}$ and $\Delta_{A,f}$ are the coupling shifts given as follows
\bea
\Delta_{\nu_L}&\simeq&\frac{t_N^2-\epsilon^2t_W^2}{16t_N^2}\frac{v^2}{\Lambda^2},\\
\Delta_{V,f}&\simeq&\frac{2\left[t_ND(f)-\epsilon t_WQ(f)\right]-T_3(f)(t_N-\epsilon t_W)}{T_3(f)-2Q(f)s_W^2}\frac{t_N-\epsilon t_W}{16t_N^2}\frac{v^2}{\Lambda^2},\\
\Delta_{A,f}&\simeq&-\frac{(t_N-\epsilon t_W)^2}{16t_N^2}\frac{v^2}{\Lambda^2}.
\eea

Using this Lagrangian, one can write the total $Z$ decay width predicted by the model,
\bea
\Gamma_{Z}=\Gamma^{\text{SM}}_{Z}+\Delta\Gamma_{{Z}},
\eea
where $\Gamma^{\text{SM}}_{Z}$ is the standard model value and the shift $\Delta\Gamma_{{Z}}$ is given by
\bea
\Delta\Gamma_{{Z}}&\simeq&\frac{m^{\text{SM}}_Z}{6\pi}\left(\frac{g}{2c_W}\right)^2\left\{\sum_fN_C(f)\left[\left(g_{0V}^Z(f)\right)^2\Delta_{V,f}+\left(g_{0A}^Z(f)\right)^2\Delta_{A,f}\right]+\frac{3\Delta_{\nu_L}}{2}\right\}\crn
&&+\frac{\Delta m_Z}{12\pi}\left(\frac{g}{2c_W}\right)^2\left\{\sum_fN_C(f)\left[\left(g_{0V}^Z(f)\right)^2+\left(g_{0A}^Z(f)\right)^2\right]+\frac{3}{2}\right\},
\eea
where $m^{\text{SM}}_Z$ is the standard model value of the $Z$ gauge boson mass, $N_C(f)$ is the color number of the fermion $f$, the sum is taken over the standard model charged fermions, and the mass shift of the gauge boson $Z$ is given by
\bea
\Delta m_Z\simeq-\frac{g}{2c_W}\frac{(t_N-\epsilon t_W)^2}{32t_N^2}\frac{v^3}{\Lambda^2}.
\eea

Note that if kinetically allowed, the gauge boson $Z$ can decay into the dark matter candidate pairs $\bar{\xi}\xi$ and $\eta^*\eta$ but these two-body decays are highly suppressed by $v^4/\Lambda^4$. From the experimental and theoretical values of $\Gamma_Z$ as aforementioned, we require $|\Delta\Gamma_Z|<0.0041$ GeV, which leads to the following bound
\bea
\Lambda\gtrsim1.14\times\frac{\sqrt{|(t_N-1.62\epsilon)(t_N-0.55\epsilon)|}}{t_N}\ \  \text{TeV}.
\eea

In the usual $U(1)_{B-L}$ gauge theory, there is no mixing of $Z$ and $Z'$, except for a contribution caused by the kinetic mixing. If $\ep$ is small enough, the total $Z$ decay width deviation is negligible, hence there is no lower bound appled for $\La$ in this case. However, in the current model, we derive $\La\gtrsim 1.14$ TeV for $|\ep|\ll1$, which is due to the dark charge breaking and quite comparable to the bounds from the invisible $Z$ decay and the Cs parity violation.

\section{\label{dart1}Collider bounds} 
\subsection{LEPII constraint}

The on-shell new gauge boson $Z'$ would not be produced at the existing $e^+e^-$ colliders if its mass is in the TeV region or higher. But, below the resonance, $Z'$ would manifestly contribute to the viable observables that make them deviating from the standard model predictions. Hence, the new gauge boson $Z'$ can be indirectly searched at the LEPII experiment through the processes $e^+e^-\rightarrow\bar{f}f$ with $f=e,\mu,\tau$. 

The processes under consideration that are induced by the exchange of the new gauge boson $Z'$ can be described by the following effective Lagrangian,
\bea
\mathcal{L}_{\text{eff}}&=&\frac{1}{1+\delta_{ef}}\left(\frac{g}{2c_W m_{Z'}}\right)^2\bar{e}\gamma_\mu [g_V^{Z'}(e)-g_A^{Z'}(e)\ga_5] e \bar{f}\gamma^\mu [g_V^{Z'}(f)-g_A^{Z'}(f)\ga_5]f,
\eea
where $\delta_{ef}=1(0)$ for $f=e$ ($f\neq e$). 

By using the relevant data of the LEPII experiment \cite{Schael:2013ita}, we impose the constraint,
\bea
\frac{4\sqrt{\pi}c_W m_{Z'}}{g\sqrt{\left[g_V^{Z'}(e)\right]^2+\left[g_A^{Z'}(e)\right]^2}}\gtrsim24.6\ \ \text{TeV},
\eea
which leads to
\bea
\Lambda\gtrsim1.23\times\frac{\sqrt{(t_N+\epsilon t_W)^2+4\epsilon^2 t_W^2}}{t_N}\ \ \text{TeV},
\eea expanded up to $(v/\La)^2$ corrections.

At the limit $|\al|\sim (v/\La)^2\ll 1$ and $|\ep|\ll1$, the new gauge boson $Z'$ couples only to left-handed charged leptons with hyperdark charge $N=1/2$, such that $g^{Z'}_V(e)\simeq g^{Z'}_A(e)\simeq \fr 1 2 c_W t_N$. Hence, it translates to a LEPII bound $\La\gtrsim 1.23$ TeV, which also agrees with the limits given above, except for the $\rho$-parameter. However, since in the $U(1)_{B-L}$ gauge theory the relevant new gauge boson couples to charged leptons by a charge $|B-L|=1$ bigger than the current case $N=1/2$, the $B-L$ breaking scale is two times bigger than our bound.

\subsection{LHC dilepton constraint}

Analogous to the LEPII, because the new gauge boson $Z'$ possesses the chiral gauge couplings to ordinary fermions with strengths different from those in the usual $U(1)_{B-L}$ theory, the $Z'$ signal strength at the LHC---which translates to a lower limit on the new physics scale for negative search result---would be dramatically changed, compared to the conventional bounds in the $U(1)_{B-L}$ theory. 

Since the LHC is energetic enough to probe $Z'$ events on shell as well as various $Z'$ couplings, in this search we appropriately take both the mixing effects coming from dark charge breaking and kinetic mixing into account and include the above constraints when turning on contribution of the mixing parameters, for a comparison at the end.

The new gauge boson $Z'$ can be resonantly produced at the LHC via the quark fusion $\bar{q}q\rightarrow Z'$ and it would subsequently decay into the standard model fermions as well as the exotic particles such as the dark matter candidate $\xi(\eta)$. The most significant decay channel of $Z'$ is given by $Z'\rightarrow l^+l^-$ with $l=e,\mu$, which has well-understood backgrounds and measures a $Z'$ that owns both couplings to quarks and leptons.    

The cross-section for this process is approximately computed in the case of the very narrow $Z'$ decay width as
\bea
\sigma(pp\rightarrow Z'\rightarrow l^+l^-) &\simeq& \frac{\pi}{3}\left(\frac{g}{2c_W}\right)^2\sum_{q}L_{q\bar{q}}(m^2_{Z'})\left\{\left[g_V^{Z'}(q)\right]^2+\left[g_A^{Z'}(q)\right]^2\right\}\crn
& &\times \frac{\Gamma(Z'\rightarrow l^+l^-)}{\Gamma_{Z'}},
\eea
where the parton luminosity $L_{q\bar{q}}$ is given by
\begin{eqnarray}
L_{q\bar{q}}(m^2_{Z'})=\int^1_{\frac{m^2_{Z'}}{s}}\frac{dx}{xs}\left[f_q(x,m^2_{Z'})f_{\bar{q}}\left(\frac{m^2_{Z'}}{xs},m^2_{Z'}\right)+f_q\left(\frac{m^2_{Z'}}{xs},m^2_{Z'}\right)f_{\bar{q}}(x,m^2_{Z'})\right],
\end{eqnarray}
where $\sqrt{s}$ is the collider center-of-mass energy, and $f_{q(\bar{q})}(x,m^2_{Z'})$ is the parton distribution function of the quark $q$ (antiquark $\bar{q}$), evaluated at the scale $m_{Z'}$. Additionally, the total $Z'$ decay width reads
\bea
\Gamma_{Z'}&\simeq&\frac{m_{Z'}}{12\pi}\left(\frac{g}{2c_W}\right)^2\sum_fN_C(f)\left\{\left[g_V^{Z'}(f)\right]^2+\left[g_A^{Z'}(f)\right]^2\right\}\crn
&&+\frac{m_{Z'}}{24\pi}\left(\frac{gt_N}{\sqrt{1-\ep^2}}\right)^2\sum^3_{i=1}\left(1-\frac{4M^2_i}{m^2_{Z'}}\right)^{3/2}\theta\left(\frac{m_{Z'}}{2}-M_i\right)\crn
&&+\frac{m_{Z'}}{48\pi}\left[\frac{g(2r-1)t_N}{\sqrt{1-\ep^2}}\right]^2\left(1-\frac{4m^2_{\eta}}{m^2_{Z'}}\right)^{3/2}\theta\left(\frac{m_{Z'}}{2}-m_{\eta}\right),
\eea
where $f$ refers to the standard model fermions and the dark fermion $\xi$ that is assumed to be radically lighter than $Z'$. The $\theta(x)$ is the step function, and $M_i$ is the $\nu_{iR}$ mass.

\begin{figure}[t]
\includegraphics[scale=0.4]{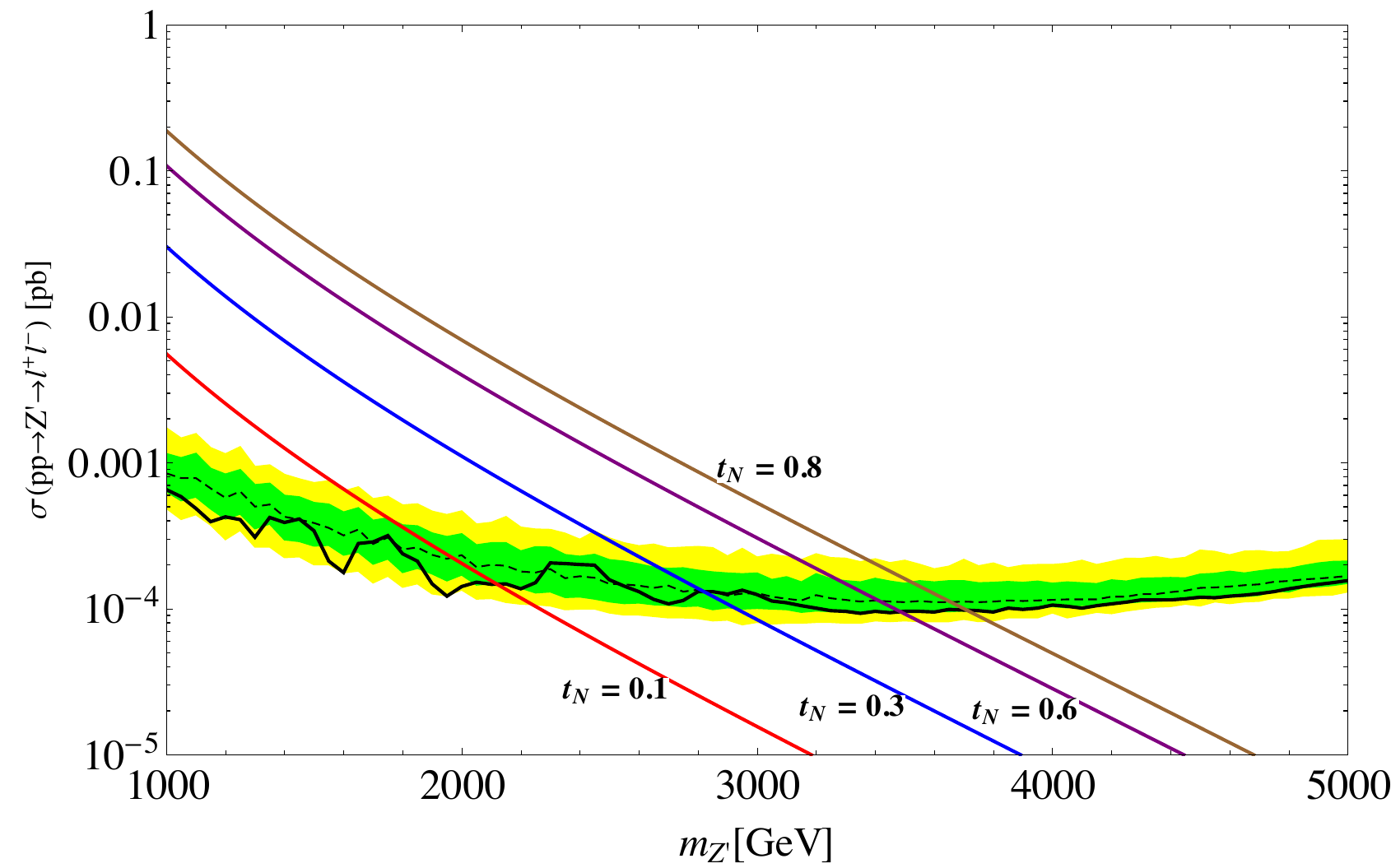}\\
\vspace{0.3cm}
\includegraphics[scale=0.4]{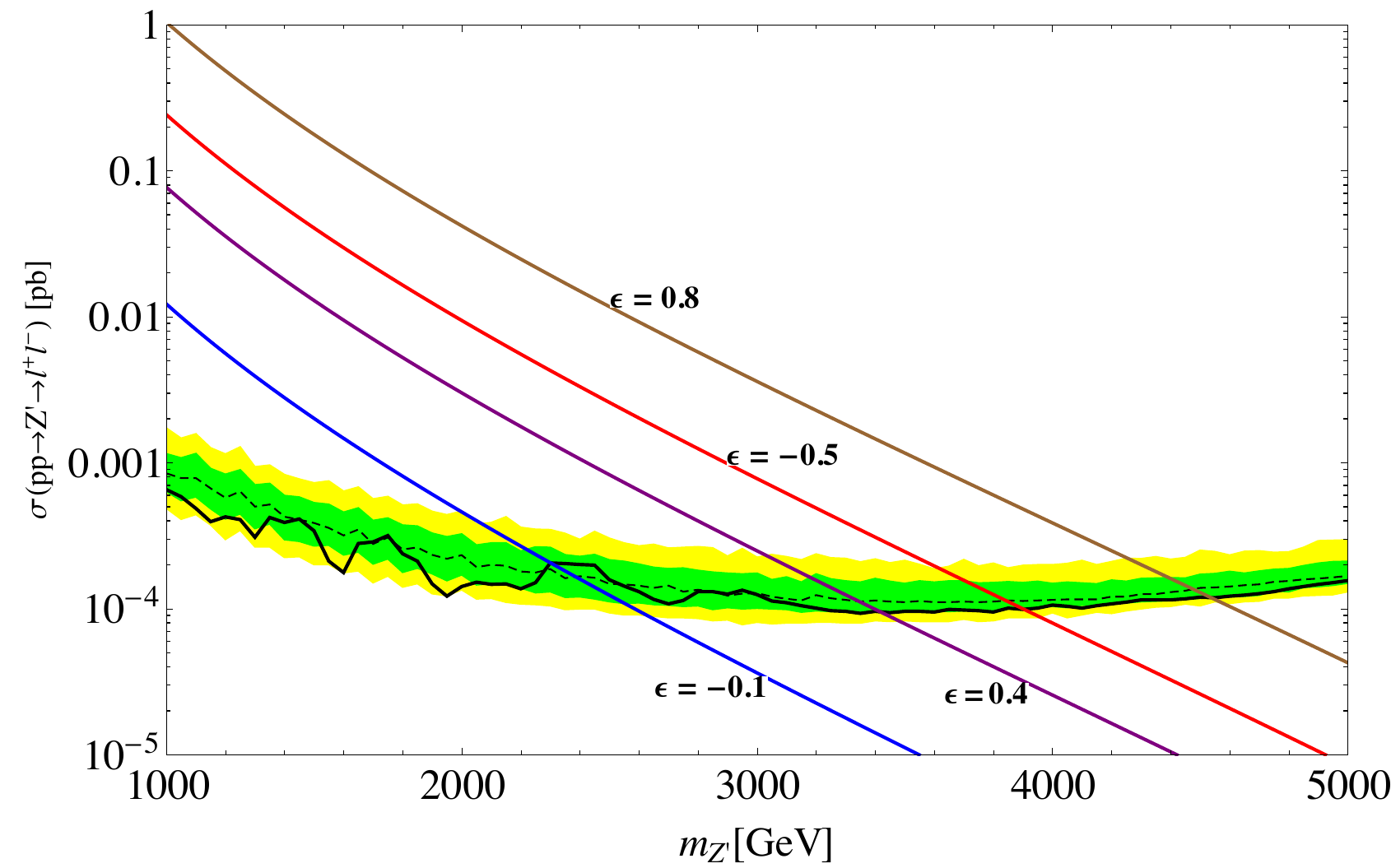}
 \caption[]{\label{dilepton-cross-sect} The cross-section for the process $pp\rightarrow Z'\rightarrow l^+l^-$ plotted as a function of the $Z'$ boson mass according to the choices of $(t_N,\ep)$, where the top and bottom panels correspond to $\epsilon=0.1$ and $t_N=0.2$, respectively. The solid and dashed black curves refer to the observed and expected limits, while the green and yellow bands refer to $1\sigma$ and $2\sigma$ expected limit, respectively \cite{Aaboud:2017buh}.}
\end{figure}

In Figure \ref{dilepton-cross-sect}, we show the dilepton production cross-section $\sigma(pp\rightarrow Z' \rightarrow l^+l^-)$ as a function of the new gauge boson mass, $m_{Z'}$, for various values of $t_N$ and $\epsilon$, with $r=1$, $M_1=M_2=M_3=m_{Z'} /3$ and $m_\eta=m_{Z'} /4$. In addition, we include the upper limits on the cross-section of this process at $95$\% credibility level using $36.1$ fb$^{-1}$ of $pp$ colision at $\sqrt{s}=13$ TeV by the ATLAS experiment \cite{Aaboud:2017buh}. In the top panel, the lower bounds on the new gauge boson mass are determined as $m_{Z'}=2.1$, $2.8$, $3.5$, and $3.7$ TeV according to $t_N=0.1$, $0.3$, $0.6$, and $0.8$, respectively, for $\ep=0.1$. Whereas, in the bottom panel, the lower bounds are
$m_{Z'}=3.9$, $2.3$, $3.4$, and $4.5$ TeV according to $\epsilon=-0.5$, $-0.1$, $0.4$, and $0.8$, respectively, for $t_N=0.2$.

\begin{figure}[h]
\includegraphics[scale=0.45]{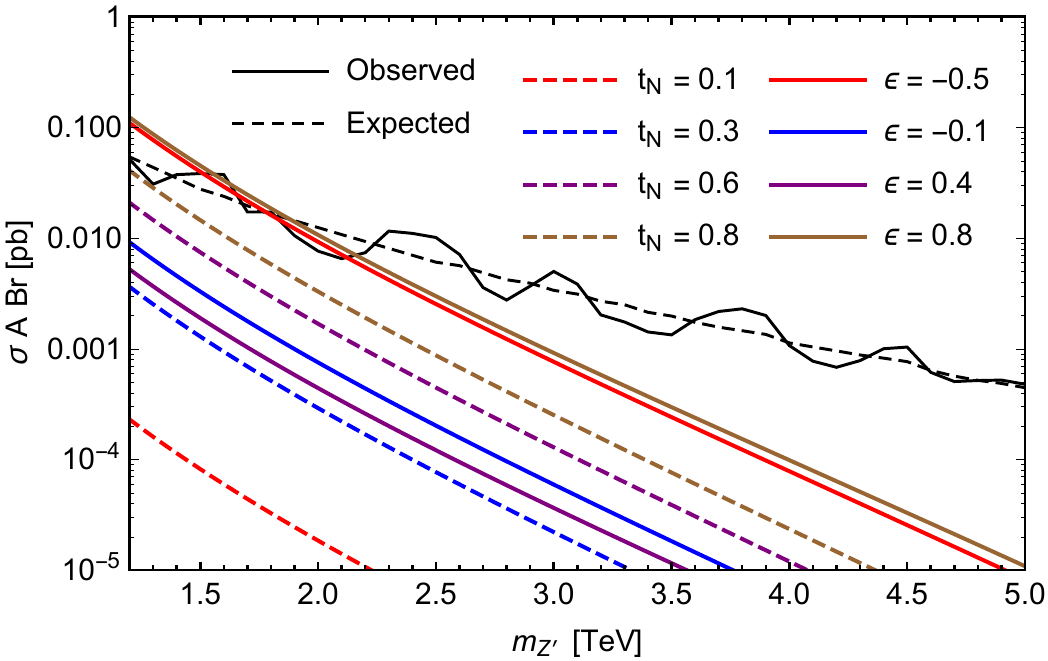}
\includegraphics[scale=0.45]{lhcbound3a}
 \caption[]{\label{dijet} The cross-section ($\sigma$) times kinematic acceptance ($A\simeq 0.4$) times branching ratio (Br) into two quarks (up quarks according to the left panel and down quarks according to the right panel) as a function of the $Z'$ boson mass ($m_{Z'}$). The solid and dashed black curves refer to the observed and expected limits, respectively \cite{ATLAS:2019fgd}, while the remaining curves are predicted by our model, in which the dashed color curves fix $\ep=0.1$, while the solid color curves fix $t_N=0.2$.}
\end{figure}

It is noteworthy that the $Z'$ boson decays not only to the leptons but also to the quarks, and thus the dijet signal can provide a lower exclusion limit for the $Z'$ mass. However, since the coupling strengths between $Z'$ and the charged leptons are approximately equal to those of $Z'$ with the quarks, as well as that the current bound on dijet signals is less sensitive than the dilepton one \cite{ATLAS:2017eqx,ATLAS:2019fgd}, the lower limit implied by the dijet search is quite smaller than that obtained from the dilepton, as explicitly shown in Figure \ref{dijet}. Hence, in the present model, the dijet bounds for the $Z'$ mass are not significant.

\begin{figure}[h]
\includegraphics[scale=0.45]{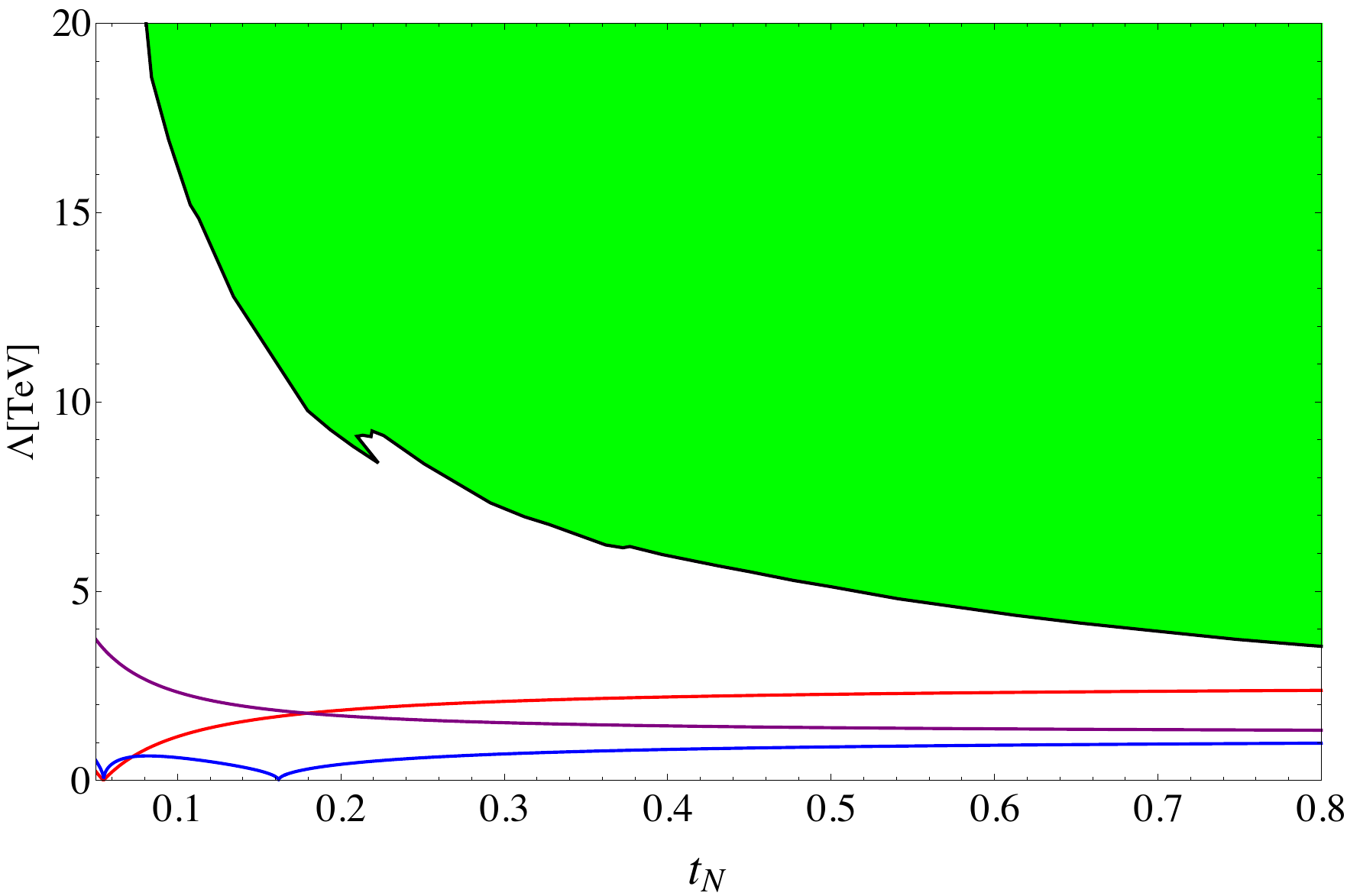}\\
\vspace{0.5cm}
\includegraphics[scale=0.3]{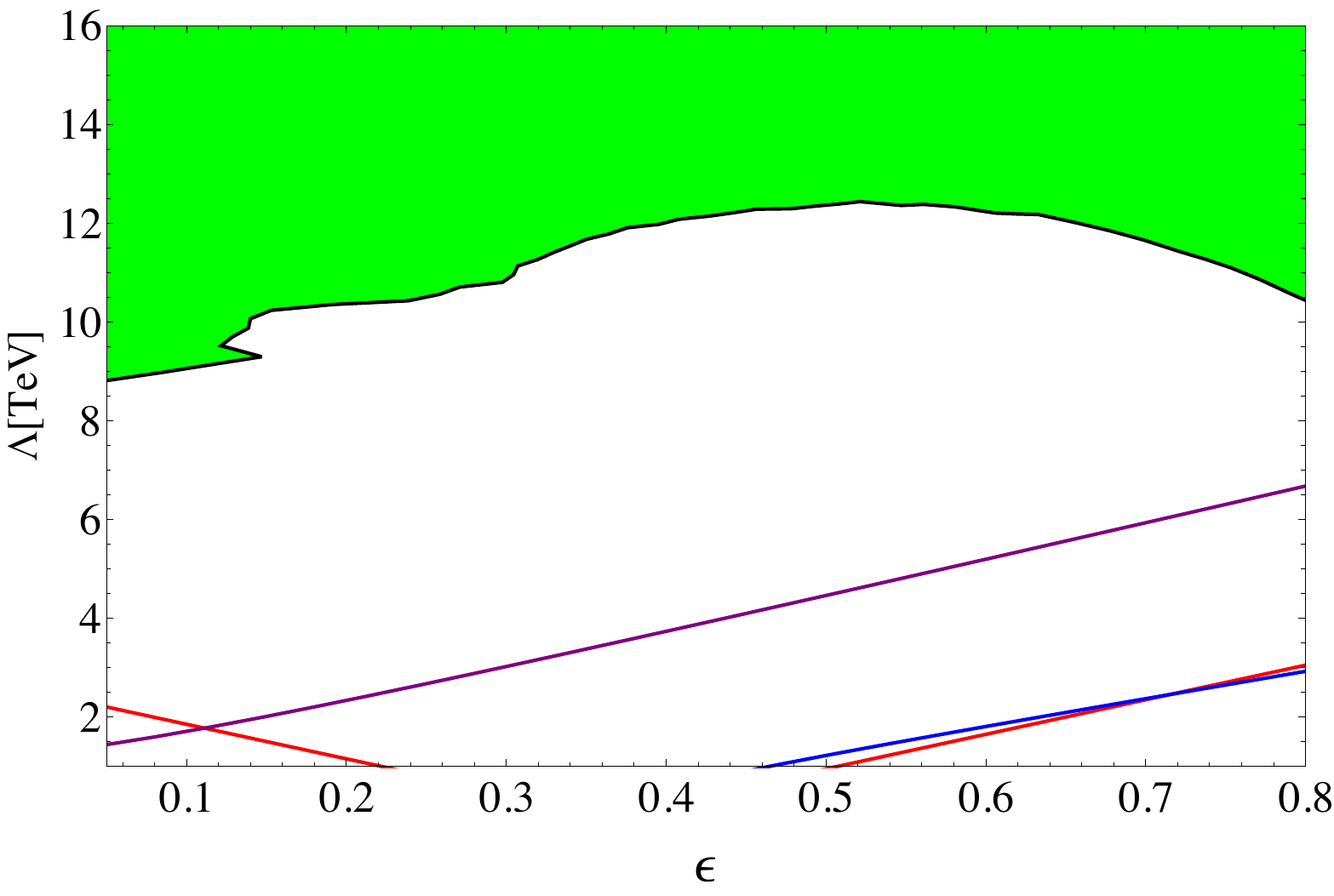}
\includegraphics[scale=0.3]{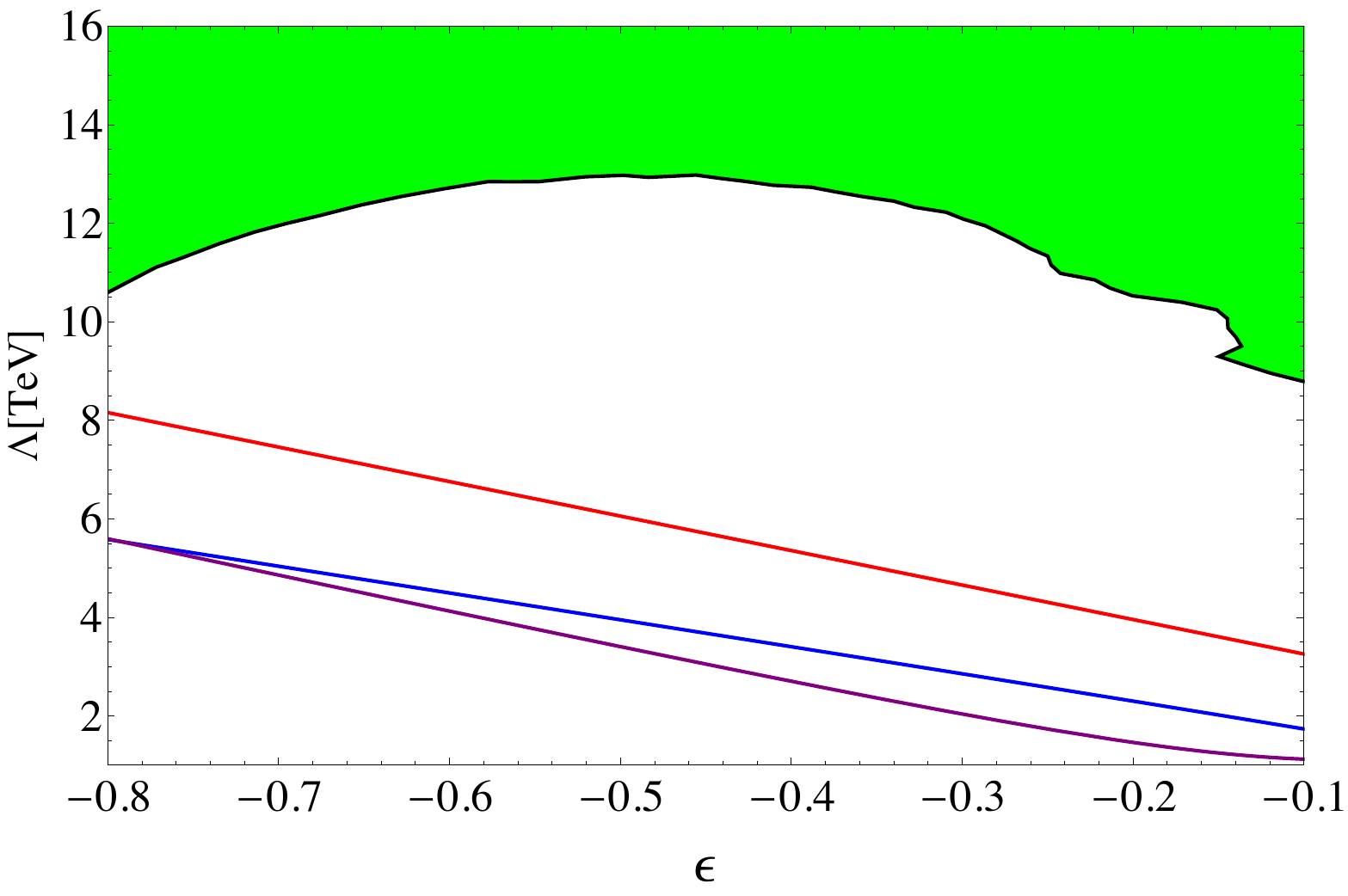}
  \caption[]{The black, blue, red, and purple curves correspond to the lower bounds obtained from the current LHC limits of the dilepton production, the precision measurement of the $Z$ decay width, the $\rho$-parameter, and the LEPII constraint, respectively. The regions that are below each of these curves are excluded; hence, the allowed parameter space is determined by the green regions. The top panel corresponds to $\epsilon=0.1$, while the bottom-left and -right panels correspond to $t_N=0.2$.}\label{APS}
\end{figure}

In Figure \ref{APS}, we combine the lower bounds, which are obtained from the current LHC limits of the dilepton production, the $\rho$-parameter, the precision measurement of the $Z$ decay width, and the LEPII constraint, to find the allowed parameter space in the $t_N$--$\Lambda$ and $\epsilon$--$\Lambda$ planes.
The top panel of this figure indicates that with $\epsilon=0.1$ the current LHC limits of the dilepton production impose the most stringent bound on the new physics scale $\La$ for the range of $t_N$ values under investigation. Similarly, the bottom-left and -right panels suggest that with $t_N=0.2$ the current LHC limits of the dilepton production impose the most stringent bound for the whole region of $\epsilon$ under consideration.

\section{\label{dmnb}Dark matter phenomenology}

As shown in Section \ref{prop}, the dark matter candidate, i.e. the lightest field of $\xi$ and $\eta$, can take an arbitrary mass. This is opposite to a previous study by one of us which limits the dark matter mass below the electron mass \cite{VanDong:2020cjf}. Hence, in this work, we have compelling scenarios that explain the dark matter abundance. 

When the dark matter candidate is significantly coupled to the normal matter in the thermal bath of the universe, the freeze-out mechanism works that determines not only the dark matter relic density but also the dark matter nature to be a weakly-interacting massive particle (WIMP). If this WIMP is sufficiently light, it may modify the synthesis of the primordial light elements of the universe, hence this scheme requires a dark matter mass to be bigger than the BBN and CMB bounds, roundly equal to the electron mass \cite{Sabti:2019mhn}. 

When the dark matter candidate is very weakly coupled to the normal matter, its annihilation rate into normal matter is always smaller than the Hubble rate and that the dark matter is not constrained by the BBN and CMB bounds. In this case, we have two folds of dark matter production, given upon the new physics scale. If the new physics scale is at TeV regime similar to the above WIMP case, a thermal freeze-in mechanism works that implies the relic density through the right-handed neutrino decay, $\nu_R\to \xi_L \eta$. Alternatively, if the new physics scale is very large, the dark matter may be asymmetrically produced from the CP-violation decay of the right-handed neutrino $\nu_R\to \xi_L\eta$, through a mechanism similar to the leptogenesis for lepton asymmetry generation. 

Let us remind that comparing the predicted neutrino masses in (\ref{nmdd}) with the neutrino oscillation data \cite{Tanabashi:2018oca} yields $\La\sim [(h^\nu)^2/f^\nu]\times 10^{14}\ \mathrm{GeV}$. Depending on the Yukawa couplings, this leads to two regimes for the seesaw scale $\La$ that is at TeV and GUT scales, respectively. Such new physics regimes are appropriate to the mentioned mechanisms for dark matter generation. The discussion delivered here updates and extends what done in \cite{VanDong:2020cjf}. 

\subsection{\label{wimp} TeV seesaw scale: WIMP dark matter} 

The seesaw scale $\La$ is in TeV regime, if $(h^\nu)^2/f^\nu$ is appropriately small, e.g. $f^\nu \sim 1$ and $h^\nu$ is similar to charged lepton Yukawa couplings. In this case, the new gauge boson $Z'$ may pick up a mass at TeV compatible to the precision test and colliders, as given above. 

Let us recall that although the charged leptons and down quarks are $P_D$ odd as the dark matter is, the dark matter cannot decay to the usual particles because of the electric and color charge conservation. In other words, the dark matter can obtain a mass larger than the usual fields, and in the early universe the dark matter can annihilate to these lighter fields, which sets the dark matter abundance by the standard thermal decoupling limit. 

We will study two scenarios where the WIMP dark matter is either a vectorlike fermion $\xi$ by imposing $m_\xi < m_\eta$ or a complex scalar $\eta$ by assuming $m_\eta < m_\xi$. 

\subsubsection{Dark matter as a fermion $\xi$}

When the fermion $\xi$ is lighter than the scalar $\eta$, the $\xi$ is stabilized responsible for dark matter. Assume that $\xi$ has a nonzero dark charge, i.e. $r\neq 0$. Processes for fermion dark matter pair annihilation into the standard model particles (leptons, quarks, Higgs, and gauge bosons) as well as possible right-handed neutrinos proceed dominantly through the contribution of the new gauge boson $Z'$ by the $s$-channel exchange diagrams. It is straightforwardly to determine the dark matter annihilation cross-section times relative velocity, given by
\bea \langle\sigma v_{\text{rel}}\rangle_{\xi\xi^c\to\text{all}} &\simeq &  \frac{g^4[g^{Z'}_V(\xi)]^2m^2_\xi }{16\pi c^4_W(4 m_\xi^2-m^2_{Z'})^2}\left(\sum_f N_C(f) \left\{[g^{Z'}_V(f)]^2+[g^{Z'}_A(f)]^2\right\}+\frac{g^2_{Z'ZH}c^2_W}{4g^2m^2_Z}\right)\crn
&&+\frac{g^4[g^{Z'}_V(\xi)]^2(C^{Z'}_{\nu_R})^2m^2_\xi }{32\pi c^4_W(4 m_\xi^2-m^2_{Z'})^2}\sum^3_{i=1}\left(1-\frac{M_i^2}{4m^2_\xi}\right)\left(1-\frac{M_i^2}{m^2_\xi}\right)^{1/2}\theta(m_\xi-M_i),\eea
where $f$ denotes the standard model fermions, the $Z'ZH$ coupling is given by $g_{Z'ZH}\simeq g^2v(\ep t_W-t_N)/2c_W\sqrt{1-\ep^2}$, and note that the $Z'$ coupling to $\nu_{iR}$ is flavor independent. The relic abundance of the dark matter fermion is $\Om_\xi h^2\simeq 0.1 \text{ pb}/\langle\sigma v_{\text{rel}}\rangle_{\xi\xi^c\to\text{all}}$. 

\begin{figure}[h]
\includegraphics[scale=0.45]{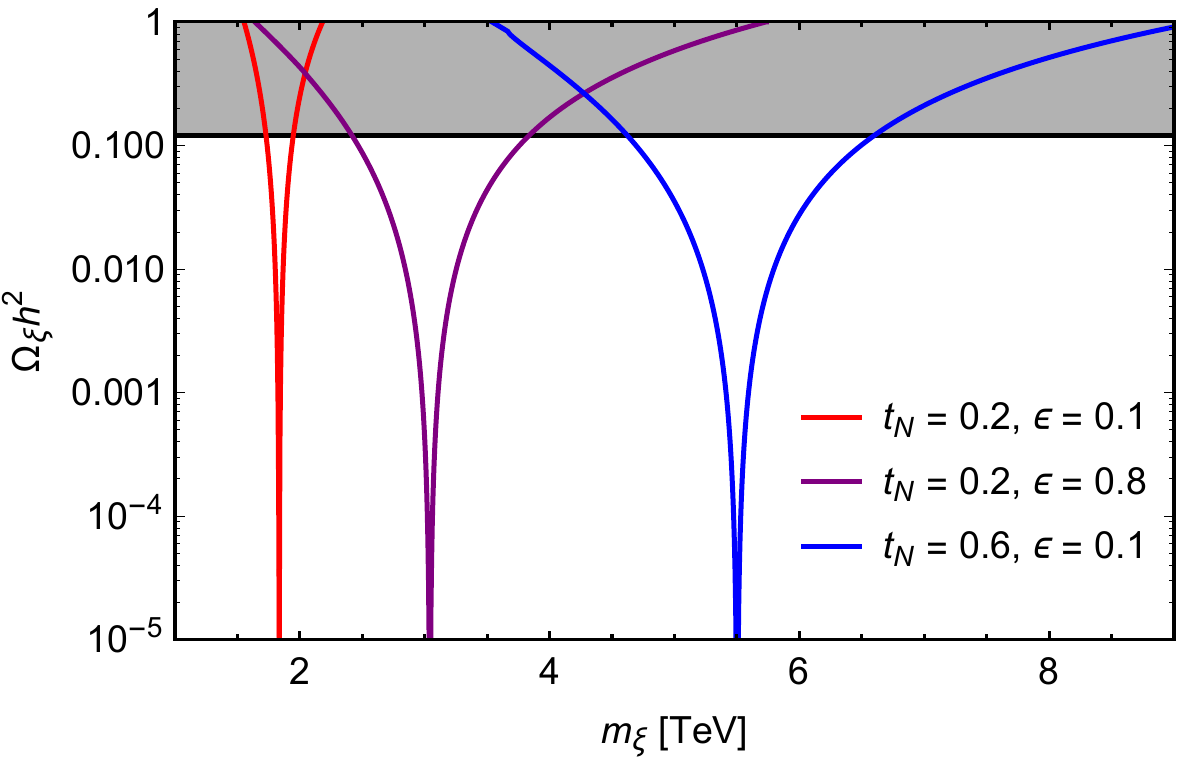}\\
\vspace{0.5cm}
\includegraphics[scale=0.4]{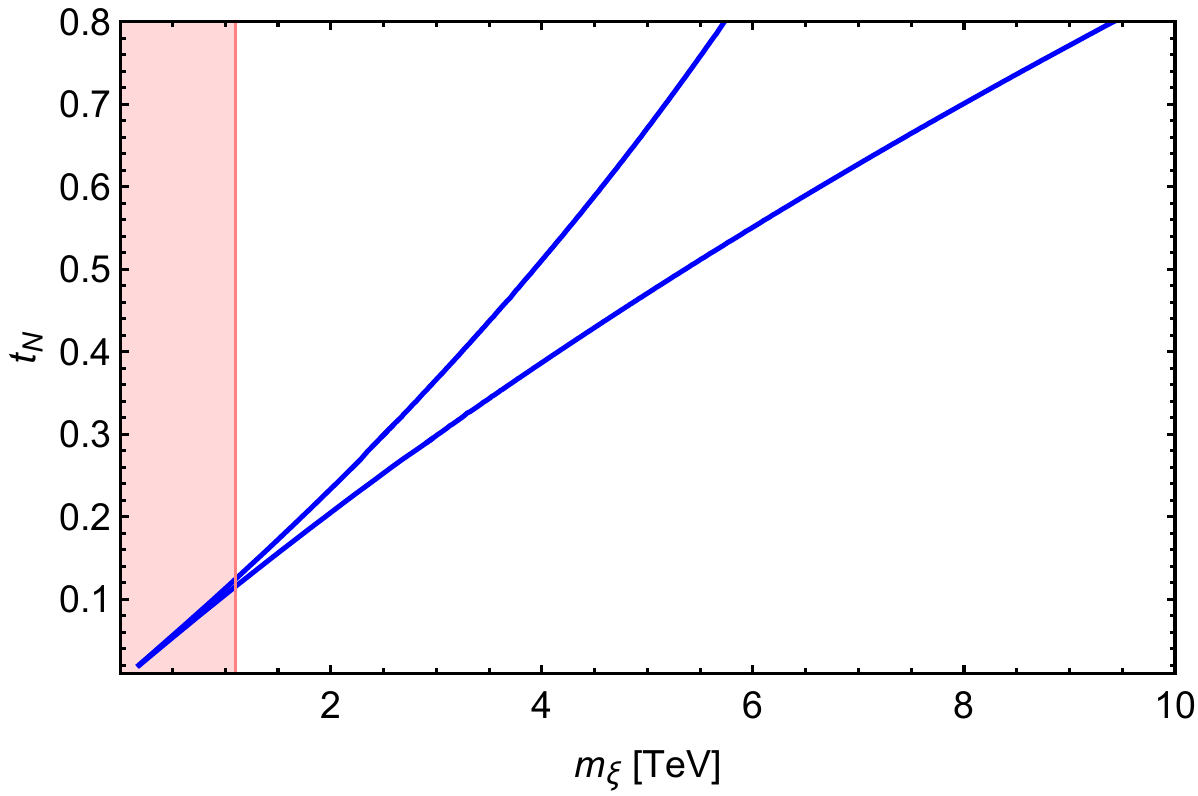}
\includegraphics[scale=0.4]{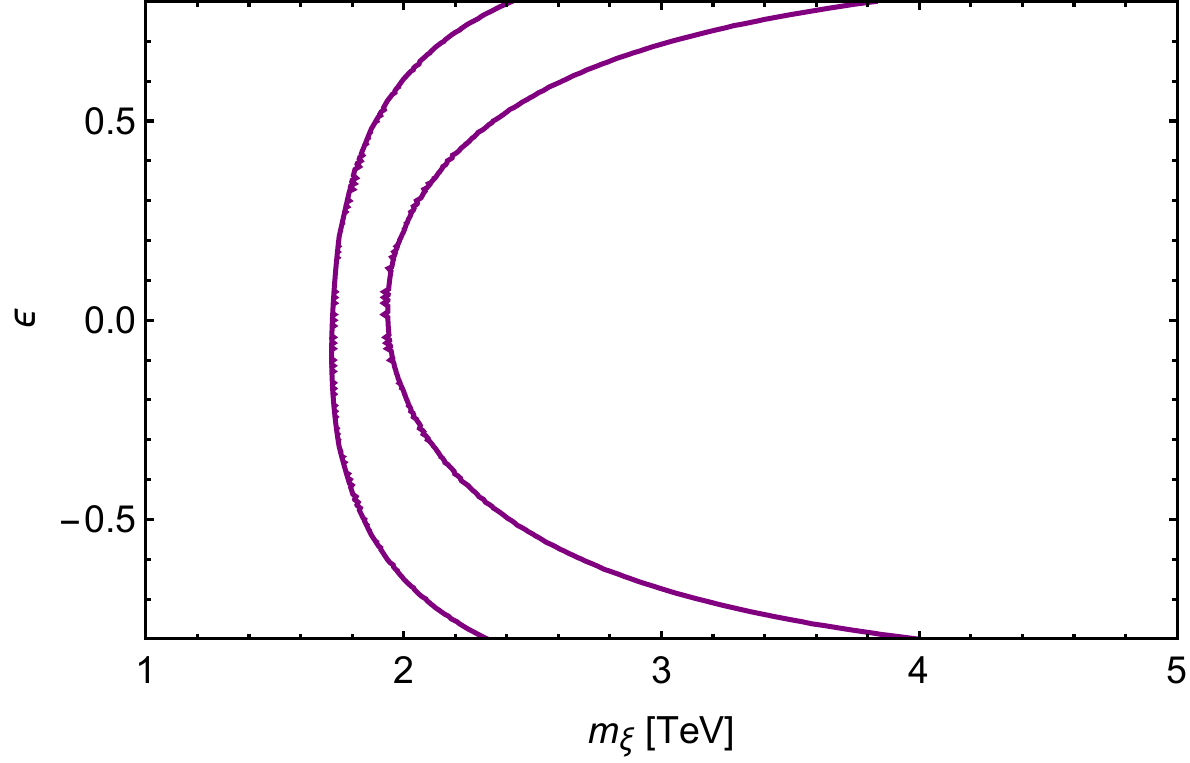}
\caption[]{Fermion dark matter relic density plotted as a function of its mass for different choices of $t_N, \ep$ (top panel), whereas in the bottom panels we contour the correct relic density according to several parameter pairs, where the bottom-left fixes $\ep=0.1$, while the bottom-right fixes $t_N=0.2$.}\label{dark1}
\end{figure}

Take $r=1$, $s^2_W=0.231$, $g=0.651$, $m_Z=91.187$ GeV, and $M_{1,2,3}=m_{Z'}/3$, as above. Let $\Lambda=14$ TeV satisfy the limits from Figure \ref{APS}, which requires $\ep=0.1$ and $0.12 \leq t_N\leq 0.8$, or alternatively $t_N=0.2$ and $-0.8 \leq \ep\leq 0.8$. In Figure \ref{dark1} top panel, we depict the relic density as a function of $m_\xi$ for the several choices of $t_N$ and $\ep$ that are viable from the mentioned regimes of Figure~\ref{APS}. Each density curve contains a resonance where the density is radically reduced, set by $m_\xi=m_{Z'}/2$. Additionally, we make contours of $\Om_\xi h^2 = 0.12$ as a function of $m_\xi$ and $t_N$ for $\ep=0.1$ in Figure \ref{dark1} bottom-left panel and as a function of $m_\xi$ and $\ep$ for $t_N=0.2$ in Figure \ref{dark1} bottom-right panel. Notice that the gray band denotes excluded parameter space according to the dark matter relic density above $0.12$ that is overpopulated, while the pink band is excluded region by the LHC for $\Lambda=14$~TeV that excludes $t_N<0.12$ or equivalently $m_\xi \approx m_{Z'}/2<1.1$ TeV as set by the resonance. From this figure, we obtain the viable dark fermion mass region to be $1.1 \text{ TeV} \leq m_\xi \leq 9.4 \text{ TeV}$ for $\ep=0.1$ and $0.12 \leq t_N \leq 0.8$, and $1.7 \text{ TeV} \leq m_\xi \leq 4 \text{ TeV}$ for $t_N=0.2$ and $-0.8 \leq \ep \leq 0.8$.

Besides providing a correct relic density, a viable dark matter candidate should evade the present constraints from  detection experiments. The strongest limits come from direct detections, which measure the spin-independent (SI) scattering cross-section of the dark matter on nucleons in target nucleus. Additionally, the scattering of the dark matter with nucleons can be described, at the microscopic level, starting from effective interactions between the dark matter and the standard model quarks. Here such interactions are dominantly contributed by $t$-channel exchange diagrams of the new gauge boson $Z'$ to be 
\be\mathcal{L}^{\text{eff}}_{\xi-\text{quark}}=(\bar{\xi}\gamma^\mu \xi)[\bar{q}\gamma_\mu(\al_qP_L+\beta_qP_R)q], \ee
where $q=u,d$, and $P_{L,R}=(1\mp\gamma_5)/2$, and
\bea\al_q &=&\frac{g^2}{4c^2_W m^2_{Z'}}g^{Z'}_V(\xi)[g^{Z'}_V(q)+g^{Z'}_A(q)],\\
\beta_q &=&\frac{g^2}{4c^2_W m^2_{Z'}}g^{Z'}_V(\xi)[g^{Z'}_V(q)-g^{Z'}_A(q)]. \eea
Hence, we obtain the SI scattering cross-section of $\xi$ on a nucleon, labelled as $\mathcal{N}\equiv p,n$ with corresponding mass $m_{\mathcal{N}}$, such as \cite{Belanger:2008sj}
\be \sigma^{\text{SI}}_\xi = \frac{4\mu^2_{\xi \mathcal{N}}}{\pi A^2}[\la_p Z+\la_n(A-Z)]^2,  \ee
in which $Z$ is the nucleus charge, $A$ is the total number of nucleons in the nucleus, $\mu_{\xi \mathcal{N}}=m_\xi m_{\mathcal{N}}/(m_\xi +m_{\mathcal{N}})\simeq m_{\mathcal{N}}$ is the reduced mass of the dark matter-nucleon system,  and
\bea\la_p &=& [2(\al_u+\beta_u)+\al_d+\beta_d]/8,\crn
 \la_n &=& [\al_u+\beta_u+2(\al_d+\beta_d)]/8 \eea
denote the effective couplings of the dark matter with protons and neutrons, respectively. 

\begin{figure}[h]
\includegraphics[scale=0.55]{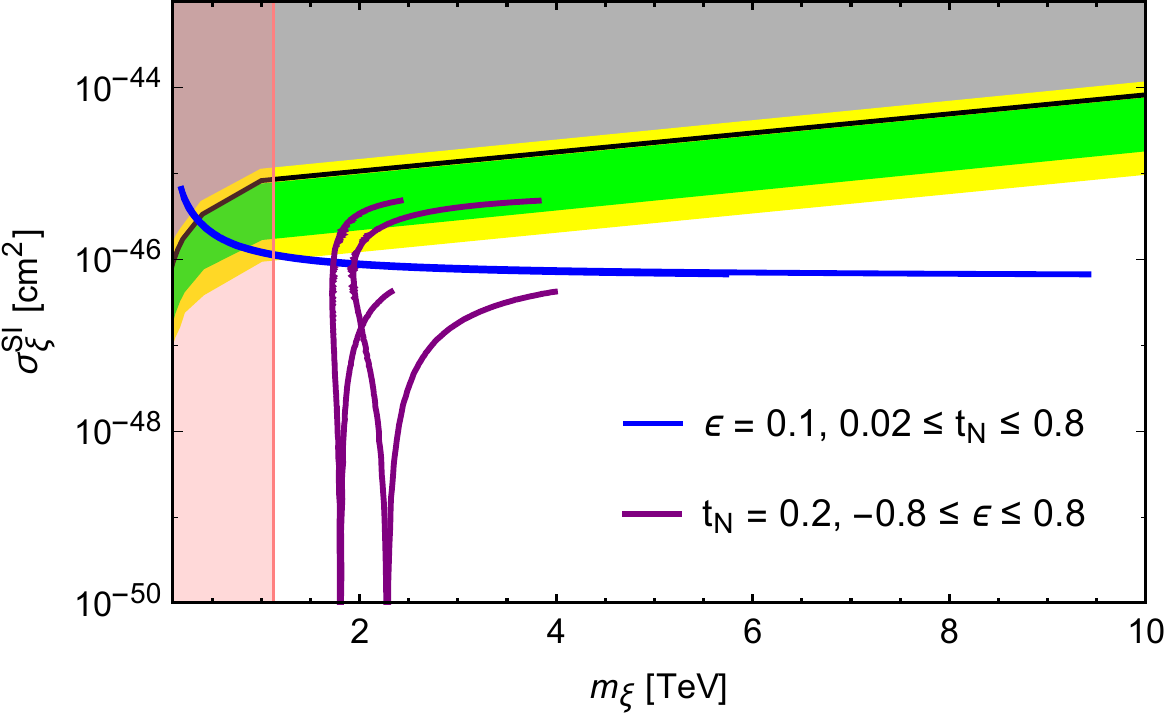}
\caption[]{The SI scattering cross-section of the dark fermion on a nucleon as a function of its mass, where the XENON1T upper limit (black line), $1\sigma$ band (green), and $2\sigma$ band (yellow), as well as the LHC (pink) and XENON1T (gray) excluded regions, are shown.}\label{SIdark1}
\end{figure}

Take $\Lambda=14$ TeV (as above), $A=131$ and $Z=54$ for the Xe nucleus, and $m_{\mathcal{N}}\simeq 1$~GeV. Assuming the correct relic density for the dark fermion, in Figure \ref{SIdark1} we plot the SI scattering cross-section as a function of the dark fermion mass according to the previously given regimes of $(\ep,t_N)$, presented as blue and purple lines, respectively. In this figure, we also include the XENON1T experimental bounds, with upper limit (black line), as well as $1\sigma$ (green) and $2\sigma$ (yellow) sensitivity bands \cite{XENON:2017vdw,XENON:2018voc}. Additionally, the pink and gray bands are the excluded regions by the LHC and the XENON1T, respectively. Note that the LHC excluded region suppresses the SI cross-section (blue line) according to $t_N<0.12$, as expected. It is clear that the viable dark fermion mass region obtained from the previous part on the relic density also satisfies the current exclusion limit of the XENON1T on direct detection.

\subsubsection{Dark matter as a scalar $\eta$}

We now consider a possibility that the complex scalar singlet $\eta$ is lighter than $\xi$ responsible for dark matter. We also assume that $\eta$ is lighter than $H'$, $Z'$, and $\nu_R$'s, for simplicity. Hence, the dark matter candidate annihilates only to the standard model particles through a contact interaction ($\la_4$) with usual Higgs fields as well as $H'$ and $Z'$ portals that set the relic density.\footnote{Since the $Z$-$Z'$ mixing angle is suppressed, i.e. $\al\sim v^2/\La^2\ll 1$, the contribution of $Z$ is small, thus omitted, similar to the fermion dark matter case.} Additionally, these interactions/portals also determine the dark matter scattering with normal matter in direct detection. The contributions of $Z',H'$ portals to the dark matter observables turn out to be quite similar to the case of the fermion dark matter with $Z'$ portal. Indeed, we find two distinct resonances in the relic density according to $m_\eta=\fr 1 2 m_{H'}$ and $m_\eta=\fr 1 2 m_{Z'}$, set by $H',Z'$ fields, respectively. However, only the $Z'$ portal governs the SI scattering cross-section, similar to the fermion dark matter case, since $H'$ does not interact with quarks at the leading order. Hence, in what follows, we will not consider the $Z',H'$ contributions. 

The most relevant phenomena are associated with the $\la_4$ coupling, such that \be V\supset\fr 1 2 \la_4\eta^*\eta(H^2+2v H),\ee which connects the scalar dark matter to the standard model particles through the usual Higgs portal. This contribution of $\la_4$ (i.e. $H$) dominates over the mentioned portals, given that $\la_4\sim 1$ is radically bigger than the gauge couplings $g,g_N$ as well as the $\la_6$ coupling that couples $\eta$ to the new Higgs $H'$. That said, the dark matter annihilation is given by the channel $\eta\eta^*\rightarrow HH$, set by the contact interaction $\la_4$, which yields the cross-section,
\bea
\langle\sigma v_{\text{rel}}\rangle_{\eta\eta^*\to HH} \simeq  \frac{7\la^2_4}{128\pi m^2_\eta}.
\eea The correct relic density, i.e. $\Om_\eta h^2\simeq 0.1\ \mathrm{pb}/\langle \sigma v_{\text{rel}} \rangle_{\eta\eta^*\to HH} \simeq 0.12$, implies a condition for the dark matter mass at TeV regime, 
\be m_\eta\simeq |\la_4|\times 2.85\text{ TeV}\sim 2.85\ \mathrm{TeV}.\label{meta} \ee

To study the dark matter direct detection, we write the effective Lagrangian that describes $\eta$-quark interactions induced by $t$-channel $H$-exchange diagrams as follows
\be \mathcal{L}^{\text{eff}}_{\eta\text{-quark}}= \frac{\la_4 m_q}{m^2_H}\eta^*\eta \bar{q}q,\ee
where $q$ denotes ordinary quarks. The SI scattering cross-section of $\eta$ on a nucleon is  \cite{Barger:2008qd}
\be\sigma^{\text{SI}}_\eta =\left(\fr{\la_4}{2\sqrt{\pi}}\frac{\mu_{\eta \mathcal{N}}}{m^2_H}\frac{m_\mathcal{N}}{m_\eta}C_{\mathcal{N}}\right)^2,  \ee
where $\mu_{\eta \mathcal{N}}=m_\eta m_{\mathcal{N}}/(m_\eta+ m_{\mathcal{N}})\simeq m_{\mathcal{N}}$, and
\be C_{\mathcal{N}}=\frac{2}{9}+\frac{1}{A}\sum_{q=u,d,s}\left[\left(Z-\frac{2}{9}A\right) f^p_{q}+(A-Z)f^n_{q}\right], \ee
in which $f^{p(n)}_q$ take the values  \cite{Hoferichter:2015dsa}
\be f^{p(n)}_{u} \simeq 0.0208 (0.0189), \hs f^{p(n)}_{d}\simeq 0.0411 (0.0451),\hs f^{p(n)}_{s} \simeq  0.043 (0.043). \ee
We estimate
\be\sigma^{\text{SI}}_\eta\simeq 1.115\times 10^{-45} \left(\frac{|\la_4|\times 2.85\text{ TeV}}{m_\eta}\right)^{2} \text{ cm}^2.\label{etaSI}\ee
Taking the result (\ref{meta}) for the correct abundance, the model predicts $\sigma^{\text{SI}}_\eta\simeq 1.115\times 10^{-45}\ \mathrm{cm}^2$, in good agreement with the XENON1T experiment for a dark matter mass at TeV regime, $m_\eta\sim 2.85$ TeV, since $\la_4\sim 1$ \cite{XENON:2017vdw,XENON:2018voc}.

\subsection{TeV seesaw scale: Freeze-in dark matter}

What happens if the dark fermion, $\xi$, has a vanished dark charge, $r=0$? [Note that the dark scalar, $\eta$, always has a nonzero dark charge, which does not play such role instead.] It is indeed a sterile particle, $\xi\sim (1,1,0,0)$, not interacting with the normal fields. It has only a coupling to the second dark field, $y\bar{\xi}_L \eta \nu_R$. We further impose $m_\xi < m_{\eta}$, so the field $\xi$ is stabilized. If $y$ is very small, the dark matter $\xi$ is very weakly coupled to the thermal bath of the universe.\footnote{Opposite to the following asymmetric dark matter, there the coupling strength between the dark fields and the normal fields is highly suppressed by the heavy $U(1)_N$ sector.} Furthermore, it is noted that since $\eta$ and $\nu_R$ are coupled to the Higgs and gauge portals via the couplings $\la_{4,6}$ and/or $g_N$, the fields $\eta,\nu_R$ are always in thermal equilibrium with the standard model plasma, as maintained by the forward/backward reactions $\mathrm{SM}+ \mathrm{SM}\stackrel{Z',H'}{\longleftrightarrow} \nu_R\nu_R (\eta\eta^*)$, in contrast to $\xi$.

Since $\xi$ does not couple to the inflation field like the one proposed below, it should have a vanished initial density. Hence $\xi$ may be later produced by the decay $\nu_R\rightarrow \xi_L \eta$ via the freeze-in mechanism, given that $m_{\nu_R}>m_{\xi}+m_{\eta}$ \cite{Hall:2009bx}. We have assumed $\nu_R$ to be the lightest among the three right-handed neutrinos, with the corresponding coupling $y$ to the dark fields. This way of dark matter genesis is opposite to the WIMP scheme studied above.  

When the cosmic temperature drops below the right-handed neutrino mass, the freeze-in mechanism supplies a relic density proportional to the decay rate $\Ga(\nu_R\to \xi_L\eta)$ \cite{Hall:2009bx}
\be \Om_{\xi}h^2\sim 0.1\left(\fr{y}{10^{-9}}\right)^2 \left(\fr{1\ \mathrm{TeV}}{m_{\nu_{R}}}\right)\left(\fr{m_{\xi}}{33.3\ \mathrm{MeV}}\right).\ee This yields a dark matter mass about $m_\xi\sim 33.3$ MeV, given that $y\sim 10^{-9}$ and the smallest right-handed neutrino mass $m_{\nu_R}\sim 1$ TeV. Of course, this dark matter mass depends on the $y$ coupling and the $\nu_R$ mass (cf. \cite{VanDong:2020cjf} for an alternative interpretation).

The possibility of a feebly-interacting massive particle, $\xi$, is very special, only for $r=0$, a tiny $y$, and $m_\xi<m_\eta$. For the case $m_\xi>m_\eta$ by contrast, the field $\eta$ becomes a dark matter, but generated by a freeze-out mechanism like the previous section. Let us turn to a more generic case, in which both $\eta,\xi$ are very weakly coupled to the standard model plasma.           

\subsection{\label{warm} Large seesaw scale: Asymmetric dark matter} 

When $\La$ is very large and that $\la_4$ is very small, the dark fields $(\xi,\eta)$ are very weakly coupled to the standard model sector (even for $r\neq 0$). The dark matter may possess any mass, not bounded by the BBN and CMB. Assuming $m_\xi<m_\eta$ without loss of generality, this means that $\xi$ is stable, responsible for dark matter. The physics happens as follows. The large field for $U(1)_N$ breaking inflates the early universe, then decaying to right-handed neutrinos. These heavy neutrinos that couple to the dark fields via complex couplings $y_a\bar{\xi}_L \eta \nu_{aR}$ CP-asymmetrically decay to the dark fields, determining the dark matter density, similar to the standard leptogenesis. Let us see this in the following. 

Given that $(h^\nu)^2/f^\nu\sim 1$, we obtain $\La\sim 10^{14}$ GeV, the scale of dark charge breaking. We first argue that this scenario of dark charge can explain the cosmic inflation driven by the dark charge breaking field, $\chi$, comparable to the one for $B-L$ breaking \cite{Huong:2015dwa,Huong:2016ybt,Dong:2018aak,VanDong:2020nwb}. 

The imaginary part of the $\chi$ field, $G_{Z'}=\sqrt{2}\Im(\chi)$, is the Goldstone boson absorbed/eaten by $Z'$ through a gauge transformation, $U=e^{-iG_{Z'}/\La}$. What remains is the real part of this field, $\Phi \equiv \sqrt{2}\Re(\chi)=\sqrt{2}U\chi\simeq \La+H'$, called inflaton. It is described by a potential, 
\be V(\Phi)=\fr 1 2 \mu^2_3\Phi^2+\fr 1 4 \la_3 \Phi^4.\label{ad123} \ee This potential cannot explain the cosmic inflation \cite{Tanabashi:2018oca}. Even if one includes Coleman-Weinberg contributions due to the couplings of $\Phi$ to $\nu_R$, $Z'$, $\phi$, and $\eta$ \cite{Coleman:1973jx}, the effective potential merely mimics the tree-level potential for large field $\Phi>\La$, whereas it predicts a too big number of $e$-folds for small field $\Phi<\La$ \cite{Dong:2018aak}. 

The inflation issue can be solved by imposing the Higgs inflation scheme for $\Phi$ instead of the usual Higgs field \cite{Bezrukov:2007ep}. For large field $\Phi>\La$, the inflaton potential is approximated to be $V(\Phi)\simeq \fr 1 4 \la_3 \Phi^4$, which preserves a scale (or conformal) symmetry. Including a nonminimal coupling of $\Phi$ to gravity, called $\delta'$, one has a Lagrangian,
\be \mathcal{L}\supset \fr{1}{2} (m^2_P+\de' \Phi^2)R+\fr 1 2 \pa_\mu \Phi \pa^\mu \Phi -V(\Phi),\ee where $R$ is Ricci scalar, $m_P=2.4\times 10^{18}$ GeV is reduced Planck mass, and $1\ll \de' \ll (m_P/\La)^2$ for consistency. Changing to the Einstein frame $\hat{g}_{\mu\nu}=\Om^2 g_{\mu\nu}$ by a conformal transformation $\Om^2 = 1+\de' \Phi^2/m^2_P$, the Lagrangian takes the canonical form, \be \hat{\mathcal{L}}=\Om^{-4}\mathcal{L}\supset \fr 1 2 m^2_P\hat{R}+\fr 1 2 \pa_\mu \hat{\Phi} \pa^\mu \hat{\Phi}-U(\hat{\Phi}),\ee with the normalized inflaton field, $\hat{\Phi}=\sqrt{3/2} m_P \ln \Om^2$, and the resultant potential, \be U(\hat{\Phi})\equiv V/\Om^4\simeq \left(\la_3 m^4_P/4 \de'^2\right)\left[1-\exp\left(-\sqrt{2/3}\hat{\Phi}/m_P\right)\right]^2,\ee which is flat for $\hat{\Phi}\gg m_P$, as desirable.   

Let $\Phi_0$ and $\Phi_e$ be the inflaton field values at horizon exit and inflation end, respectively. The slow-roll parameters $\ep(\Phi)$, $\eta(\Phi)$, and $\zeta(\Phi)$, the curvature perturbation $\Delta^2_{\mathcal{R}}(\Phi)$, and the number of $e$-folds $N(\Phi)$ can be directly deduced from $U(\hat{\Phi})$. The inflation ends at $\ep(\Phi_e) \simeq 1$, giving $ \Phi^2_e\simeq (2/\sqrt{3}\de')m^2_P$. The standard cosmology \cite{Tanabashi:2018oca} yields both $N(\Phi_0)\simeq 60$, implying $ \Phi^2_0\simeq (84.84/\de')m^2_P$, and $\Delta^2_{\mathcal{R}}(\Phi_0)=2.215\times 10^{-9}$ at pivot scale $k_0=0.05$ Mpc$^{-1}$, supplying $\de'/\sqrt{\la_3}\simeq 5.04\times 10^4$. We achieve the inflation observables at horizon exit, such as the spectral index $n_s\simeq 0.967$, the tensor-to-scalar ratio $r\simeq 0.00296$, and the running index $\al\simeq -5.23\times 10^{-4}$, in agreement to the experiments \cite{Planck:2015sxf}.

\begin{figure}[h]
\bc
\includegraphics[scale=1]{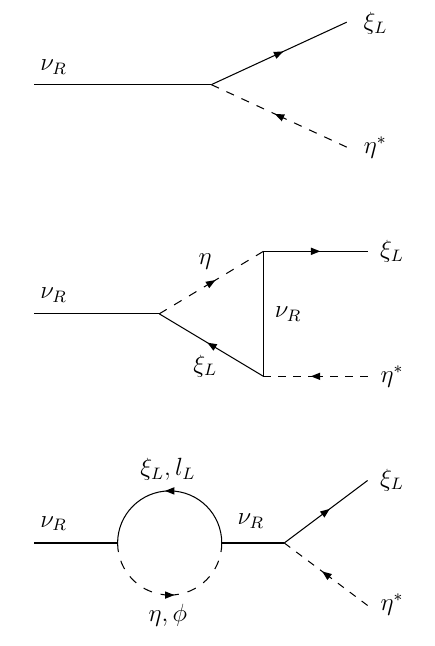}
\caption[]{\label{asymdm} CP violating decay of $\nu_R$ creating the asymmetry dark matter of the universe.}
\ec
\end{figure}

After the inflation, the right-handed neutrinos may directly be created by the inflaton decay, $\Phi\rightarrow \nu_R\nu_R$, which reheats the universe.\footnote{At the scale of the dark charge breaking, $\Om^2=1+\de' \La^2/m^2_P\simeq 1$ is close the identity, hence the fields in Einstein frame coincide with those in Jordan frame. The hat mark on fields may be omitted.} Alternatively, the fields $\nu_R$ may be created in the cosmic plasma through thermalizing other states during the preheating or/and reheating stages. These right-handed neutrinos presented in the early universe then decay to dark fields, $\nu_R\to \xi_L \eta$, through diagrams as depicted in Fig. \ref{asymdm}, which both violates CP symmetry and drops out of thermal equilibrium when the universe cools down, explaining the abundance of the asymmetric dark matter of the universe. This process is similar to the CP-violating decay of $\nu_R$ to normal matter, i.e. $\nu_R\to l_L\phi$, that explains the lepton asymmetry via leptogenesis \cite{Fukugita:1986hr}.

Assume the right-handed neutrino masses to be hierarchical, $M_1\ll M_{2,3}$. The asymmetric dark matter generation is proceeded through $\nu_{1R}$ decay, determined by
\be \ep_{\mathrm{DM}}=\fr{\Ga(\nu_{1R}\rightarrow \xi \eta)-\Ga(\nu_{1R}\rightarrow \bar{\xi} \bar{\eta})}{\Ga_{\nu_{1R}}},\ee where $\Ga_{\nu_{1R}}$ is the total width of $\nu_{1R}$ which must include $\nu_{1R}\to l_L \phi$ too. From Fig. \ref{asymdm}, it is straightforwardly to derive 
\be \ep_{\mathrm{DM}}=\fr{\sum_i \Im[y^*_i y_1(y^*_i y_1+h^{\nu *}_{ai} h^\nu_{a 1})]M_1/M_i}{8\pi (y^*_1 y_1+2 h^{\nu *}_{a1} h^\nu_{a1})}.\ee Note that the 2-to-2 scatterings between normal fields $(l\phi)$ and dark fields $(\xi \eta)$ are suppressed because they are mediated by superheavy particles $\nu_R$, $Z'$, and $H'$. Hence, the Boltzmann equation that describes the abundance yield of dark matter asymmetry, $Y_{\mathrm{DM}}$, is decoupled from that for normal matter. The abundance yield takes the form,  
\be Y_{\mathrm{DM}}=\eta_{\mathrm{DM}} \ep_{\mathrm{DM}}Y^{\mathrm{eq}}_{\nu_{1R}}(0),\ee where $Y^{\mathrm{eq}}_{\nu_{1R}}(0)=135\zeta(3)/(4\pi^4g_*)\simeq 4\times 10^{-3}$. And, the efficiency factor is given by the Boltzmann equation to be \be \eta_{\mathrm{DM}}\simeq \fr{H(T=M_1)}{\Ga(\nu_{1R}\to \xi\eta)}\simeq \fr{170 M_1}{y^*_1 y_1 m_P},\ee 
appropriate to the strong washout regime, $\eta_{\mathrm{DM}}\ll 1$, where $H=0.33\sqrt{g_*}T^2/m_P$ is the Hubble rate at the asymmetric $\nu_{1R}$ decay, $T=M_1$, with $g_*=106.75$ counting the effective number of degrees of freedom.  

The relic abundances of dark matter and normal matter have been well measured, giving a relation $\Om_{\mathrm{DM}}\simeq 5 \Om_B$. Hence, the dark matter mass obeys \be m_\xi \simeq 5m_pY_B Y^{-1}_{\mathrm{DM}} \simeq \fr{1.6\times 10^{-8}  m_p m_Py^*_1y_1(y^*_1y_1+2h^{\nu*}_{a1}h^\nu_{a1})}{M^2_1\sum_i \Im[y^*_i y_1(y^*_i y_1+h^{\nu *}_{ai} h^\nu_{a 1})]/M_i},\label{dnt201}\ee where $m_p\simeq 1$ GeV is the proton mass, and $Y_B\simeq 0.87\times 10^{-10}$ is the baryon to entropy ratio~\cite{Tanabashi:2018oca}. Expanding the inflation potential, we obtain the inflaton mass $m_{\hat{\Phi}}=\sqrt{\la_3/3}m_P/\de'\simeq 2.77\times 10^{13}$ GeV. Thus, one can take \be M_1\ll m_{\hat{\Phi}} \sim M_{2} \sim M_3,\ee such that the inflaton suitably decays to $\nu_{1R}$, not to $\nu_{2,3R}$, after the inflation end.\footnote{If $M_1$ is close to $m_{\hat{\Phi}}$, the universe undergoes a period of preheating, waiting for necessary inflaton oscillations before decay, $\hat{\Phi}\to \nu_{1R}\nu_{1R}$. In this preheating, the nonperturbative decay $\hat{\Phi}\to Z'Z'$ exists \cite{Kofman:1994rk}, whose products rapidly thermalize producing a cosmic plasma with temperature much beyond the conventional reheating temperature \cite{Chung:1998rq}. The right-handed neutrino can be created by this $Z'$ thermalization and thus populated in the early universe before the reheating, as mentioned.} Since $(h^\nu)^2\sim f^\nu\sim M/\La$ is radically smaller than 1, we assume $(h^\nu)^2\ll y^2 \sim 1$, appropriate to the strong washout regime, $\eta_{\mathrm{DM}}\ll1$. Omitting the normal field contributions by the $(h^\nu)^2$ terms in (\ref{dnt201}), as well as setting $M_2=M_3$ and $y_{2,3}=y_1 t e^{-i\kappa}$ where $t,\kappa$ are real, we have 
\be m_\xi \simeq 69 \left(\fr{10^{-1}m_{\hat{\Phi}}}{M_1}\right)^2\left(\fr{M_{2,3}}{m_{\hat{\Phi}}}\right)\left(\fr{1}{t^2\sin 2\kappa}\right) \ \mathrm{MeV}.\ee Assuming $t=|y_{2,3}|/|y_1|\sim 1$, $\kappa\sim \pi/4$, and $M_{1}\sim 10^{-1}m_{\hat{\Phi}}$, we obtain the dark matter mass $m_\xi\sim 69$ MeV. Note that $\eta$ should have a mass larger than that of the dark matter $\xi$. 

Hence, the dark matter gains a correct abundance with a mass $m_\xi \sim 69$ MeV, given that $\nu_R$'s couple to the dark sector to be stronger than to the normal sector, i.e. $|y_{2,3}|\sim |y_1| > |h^\nu|$, and that the CP-violation phase is maximal, $\mathrm{Arg}(y_1/y_{2,3})=\pi/4$. A similar process, known as the leptogenesis, also generates a baryon asymmetry, \be Y_B\simeq 1.3\times 10^{-3} \eta_{\mathrm{NM}}\ep_{\mathrm{NM}}\sim 4.4\times 10^{-10}\sum_{j=2,3}\Im(e^{i\pi/4}h^{\nu*}_{aj}h^\nu_{a1})/(h^{\nu *}_{a1}h^\nu_{a1}),\ee in order of $10^{-10}$, comparable to the observation. Since $\xi$ communicates with normal matter only through the superheavy $Z'$ portal if $\xi$ has a nonzero dark charge (otherwise, it is sterile), $\xi$ does not significantly interact with the detectors in direct detection.        
 
Last, but not least, the mass of the asymmetric dark matter as well as that of the freeze-in dark matter in the previous section are, as obtained, all beyond MeV scale. Hence, there does not exist any sub-MeV dark field. This implies that there is no new relativistic degree of freedoms present during the BBN. Additionally, the dark matter candidates in both the schemes are always stabilized for which they neither decay nor modify the predicted number density of the known matter components. That said, the asymmetric and freeze-in dark matter schemes are consistent with the cosmic observations, predicted by the standard cosmology.

\subsection{Necessity of this study of dark matter}

Dark matter is known to be electrically and color neutral. However, its stability closely related to these universal charges was not interpreted, to our best knowledge. Our basic idea is to propose a dark charge, mirror of electric charge, thus defining a hyperdark charge through the $T_3$ operator. The interest is able to couple the hyperdark charge breaking field to right-handed neutrinos, raising it to the rank of Majoron. The residual gauge symmetry works in such a way that the electric and color charge conservations are crucially to keep the dark matter stable. The dark sector states defined by the dark parity reveal two singlets, the fermion $\xi$ and the scalar $\eta$, to be the simplest candidates for dark matter. This interpretation of dark matter stability leads to a novel gauge portal that communicates dark matter to normal matter, besides yielding characteristic signatures at the precision test and colliders. 

Indeed, this construction is precisely to limit the types of charges between $Z'$ and ordinary fermions (including right-handed neutrinos as well) by anomaly cancelation, given through the $\delta$-charge relations, such as \be N(l_L,q_L,\nu_R,e_R,u_R,d_R)=-1/2+\delta,1/6-\delta/3,\delta,-1+\delta,2/3-\delta/3,-1/3-\delta/3,\ee respectively. Hence, $Z'$ has effective chiral couplings to normal matter, unlike a vectorlike $B-L$ gauge field. Besides governing the dark matter observables as presented, $Z'$ can be used for distinguishing our dark matter model in experiment. First, the scattering of dark matter with target nuclei or electrons in direct detection experiment violates the parity conservation explicitly. Second, the way that $Z'$ couples to the normal matter differently from the usual theories could sign novel ratios of dark matter annihilation in neutrino/positron/antiproton in indirection detection experiment. For the WIMP scenarios that are governed by the $Z'$ portal and received a dark matter mass in tens of GeV, we obtain \be \langle \sigma v\rangle_{\nu\nu^c}:\langle \sigma v\rangle_{e^-e^+}:\langle \sigma v\rangle_{bb^c}=(\delta^2-\delta+1/4):(2\delta^2-3\delta+5/4):(2\delta^2/3+\delta/3+5/12),\ee where the annihilation in antiproton includes the contribution of $bb^c$ channel. Whereas, when the dark matter mass is beyond the weak scale, we deduce \be \langle \sigma v \rangle_{\nu\nu^c}:\langle \sigma v \rangle_{e^-e^+}:\langle \sigma v\rangle_{tt^c}=(\delta^2-\delta+1/4):(2\delta^2-3\delta+5/4):(2\delta^2/3-5\delta/3+17/12),\ee where the annihilation in antiproton is accounted for $tt^c$ channel. Such ratios that experiments could measure allow for reconstructing the $\delta$ charge, a characteristic signature that allows the discovery of this type of model. The model under consideration with $\delta=1$ yields the annihilation ratios to be either $3:3:17$ or $3:3:5$ corresponding to the hadronic channels to be either $bb^c$ or $tt^c$, respectively.\footnote{For comparison, the annihilation ratios in neutrino/positron/antiproton in the usual $B-L$ theory are $3:6:2$, respectively, indistinguishable to up- and down-type quark channels.} Above, we have assumed the right-handed neutrinos to be heavier than the WIMPs. Also, we do not look into any indirect detection experiment in detail, a task to be taken up elsewhere. Surely, the WIMP annihilation cross-sections that obey the correct relic density, i.e. $\langle \sigma v\rangle \sim 1$ pb, would easily evade such an indirection detection constraint for an appropriate WIMP mass.         

Last, but not least, since the dark matter stability is closely related to the neutrino mass generation, the dark matter candidates may interact with right-handed neutrino portals, alternatively to the $Z'$ portal. It is interesting that the leptogenesis works, appropriately producing both normal and dark matter asymmetries, a feature not naturally imposed in the usual $B-L$ theory since $\xi$ is completely sterile and omitted. Additionally, in the other scheme, dark matter can be produced via freeze-in decay of right-handed neutrinos.         

\section{\label{conl}Conclusion}

We have proved that a dynamical dark charge naturally arises as a variant of the usual electric charge. The dark dynamics interprets the right-handed neutrinos to be fundamental fields which are both charged under the dark charge and received large Majorana masses through the dark charge breaking. This dark charge breaking implies not only the electric charge quantization as fixed by the mentioned Majorana masses, but also the observed, small neutrino masses given in terms of a canonical seesaw when the weak breaking proceeds. It is noteworthy that the dark charge breaking also supplies a residual dark parity, such as $P_D\equiv (-1)^{3D+2s}$, providing a stable dark matter candidate. This kind of dark matter stability symmetry differs from the most studies: Although the charged leptons and down quarks are odd under the dark parity as the dark matter is, the dark matter is stabilized simply by the electric and color charge conservations.  

The new physics effect comes from the $U(1)_N$ sector that determines the dark charge. We have examined the new physics contributions to the $\rho$-parameter, the total $Z$ decay width, the LEPII and LHC dilepton searches. The results indicate that the dark charge breaking scale $\La$ and the new gauge boson $Z'$ mass are bounded at several TeVs.  

Depending on the magnitude of the seesaw scale as well as the coupling strength between the normal and dark sectors, the novel scenarios for dark matter production may be recognized. The large seesaw scale scheme for the dark charge breaking generates appropriate asymmetric fermion dark matter with a dark matter mass around 69 MeV, with the suitable choice of parameters. This dark matter genesis is analogous to the lepton asymmetry production from the standard leptogenesis. Indeed, both kinds of the matter relic arise from the lightest right-handed neutrino decay. By contrast, the TeV seesaw scale scheme for the dark charge breaking is appropriate to the production of the WIMP dark matter. In the special case, the sterile fermion dark matter may be alternatively produced from a freeze-in decay of the right-handed neutrino. All such dark matter generation schemes are manifestly governed by the $U(1)_N$ gauge symmetry, i.e. the dark dynamics. 

Finally, multicomponent dark matter can be recognized due to the existence of the many solutions of $U(1)_N$ factors or $Z_2$-larger residual symmetries induced within each $U(1)_N$, a task to be conducted elsewhere \cite{VanLoi:2021dzv}.  

\section*{Acknowledgments}
This research is funded by Vietnam National Foundation for Science and Technology Development (NAFOSTED) under grant number 103.01-2019.353.

\appendix

\section{\label{alge} Current algebra approach} 

Consider the $SU(2)_L$ symmetry of weak isospin, $T_i$ ($i=1,2,3$), in which the left-handed fermions transform as isodoublets, $l_L=(\nu_L\ e_L)^T$ and $q_L=(u_L\ d_L)^T$, whereas the corresponding right-handed fermions are put in singlets, where generation indices have been suppressed. Vector-like fermion introduced later for dark matter is not counted, without loss of generality. Further, we assign the electric charge and the dark charge to each fermion, such as $Q(\nu,e,u,d)=0,-1,2/3,-1/3$ and $D(\nu,e,u,d)=\delta,\delta-1,2/3-\delta/3,-1/3-\delta/3$, respectively. The latter charge values can be extracted from (\ref{decq}).     

The covariant derivative relevant to $SU(2)_L$ is \be D_\mu=\pa_\mu + ig T_i A_{i\mu}=\pa_\mu + ig[(T_+ W^+_\mu +H.c.)+T_3 A_{3\mu}],\ee where $T_\pm\equiv (T_1\pm i T_2)/\sqrt{2}$ and $W^{\pm}\equiv (A_{1} \mp i A_{2})/\sqrt{2}$. Thus, the gauge interaction of fermion multiplets, commonly labeled as $F$'s, takes the form, \bea \mathcal{L}&\supset& \sum_F \bar{F}i\ga^\mu D_\mu F\crn
& \supset&\sum_F\left[ (-g \bar{F}_L \ga^\mu T_+ F_L W^{+}_\mu +H.c.)-g \bar{F}_L\ga^\mu T_3 F_L A_{3\mu}\right].\eea This leads to weak currents in the Lagrangian, $\mathcal{L}\supset -gJ^\mu_+ W^+_\mu +H.c.-gJ^\mu_3 A_{3\mu}$, such that \be J^\mu_\pm =\sum_F \bar{F}_L\ga^\mu T_\pm F_L,\hs J^\mu_3=\sum_F \bar{F}_L\ga^\mu T_3 F_L.\ee 

The weak currents give rise to the corresponding weak charges, 
\bea && T_+(t)\equiv \int d^3x J^0_+=\fr{1}{\sqrt{2}} \int d^3x (\nu^\dagger_{L} e_{L}+u^\dagger_{L} d_{L}),\crn
&& T_3(t)\equiv \int d^3x J^0_3=\fr 1 2 \int d^3x (\nu^\dagger_{L} \nu_{L}-e^\dagger_{L} e_{L} +u^\dagger_{L} u_{L}-d^\dagger_{L} d_{L}),
\eea and $T_-(t)=[T_+(t)]^\dagger$. Using the canonical anticommutation relation, $\{f(\vec{x},t),f^\dagger(\vec{y},t)\}=\de^{(3)}(\vec{x}-\vec{y})$, the weak charges obey the $SU(2)_L$ algebra, as expected, \be [T_+(t),T_-(t)]=T_3(t),\hs  [T_3(t),T_\pm(t)]=\pm T_\pm(t).\ee  

The $Q(t)$ and $D(t)$ charges are given by 
\bea Q(t) &=&\int d^3x F^\dagger Q F=\int d^3x \left[-e^\+_{L} e_{L}+\fr 2 3 u^\+_{L}u_{L} -\fr 1 3 d^\+_{L}d_{L}+(RR)\right],\\
D(t) &=&\int d^3x F^\dagger D F\crn
&=&\int d^3x \left[\delta \nu^\+_{L}\nu_{L}+(\delta-1)e^\+_{L} e_{L}+\fr{2-\delta}{3} u^\+_{L}u_{L} -\fr{1+\delta}{3}d^\+_{L}d_{L} +(RR)\right].
\eea $Q(t)$ and $D(t)$ are not proportional to $T_3(t)$, because they have the right currents. Hence, $Q,D$, and the weak isospin do not form a closed algebra, under which we base our theory. Further, we derive 
\be [Q(t),T_{\pm}(t)]=\pm T_\pm (t),\hs [D(t), T_{\pm}(t)]=\pm T_\pm(t),\ee implying that $Q,D$ do not commute with the weak isospin. 

We obtain  
\bea Q(t)-T_3(t)&=&\int d^3x \left[-\fr{1}{2}l^\dagger_{L} l_{L}+\fr{1}{6}q^\+_{L}q_{L}-e^\dagger_{R}e_{R}+\fr 2 3 u^\dagger_{R}u_{R}-\fr 1 3 d^\+_{R}d_{R}\right]\crn
&\equiv& \int d^3x F^\dagger Y F,\\ 
D(t)-T_3(t)&=&\int d^3x \left[\left(\delta-\fr 1 2\right)l^\dagger_{L} l_{L}+\left(\fr{1}{6}-\fr{\delta}{3}\right)q^\+_{L}q_{L}\right.\crn
&&\left.+(\delta-1)e^\dagger_{R}e_{R}+\fr{2-\delta}{3} u^\dagger_{R}u_{R}-\fr{1+\delta}{3} d^\+_{R}d_{R}\right]\crn
&\equiv& \int d^3x F^\dagger N F,\eea
which yield two new abelian charges, $Y$ and $N$, with their values for multiplets coinciding with those in the main text, respectively. 

It is easily to check that $Y(t)=\int d^3 x F^\dagger Y F$ and $N(t)=\int d^3 x F^\dagger N F$ commute with the weak isospin and are linearly independent. Hence, we conclude that the manifest gauge symmetry must be \be SU(2)_L\otimes U(1)_Y\otimes U(1)_N,\ee apart from the color group. Additionally, $Y$ and $N$ define the electric charge and the dark charge given, respectively, by \be Q-T_3= Y,\hs D-T_3=N.\ee 

Let us stress that the $SU(2)_L$ weak isospin theory contains in it two conserved and noncommutative charges, $Q$ and $D$, and that the requirement of algebraic closure between them yields the $SU(2)_L\otimes U(1)_Y\otimes U(1)_N$ gauge model, describing the electroweak and dark interactions. Interestingly, the weak and dark interactions are unified in the same manner that the electroweak theory does so for the weak and electromagnetic interactions. 

\section{\label{anomaly}Anomaly checking}
For convenience in reading, let us recall the full gauge symmetry,
\be SU(3)_C\otimes SU(2)_L \otimes U(1)_Y\otimes U(1)_N,\ee and collect the $U(1)_{Y,N}$ quantum numbers in Table~\ref{alb}.
\begin{table}[h]
\bc
\begin{tabular}{c|ccccccc}
\hline\hline
Multiplet & $l_L$ & $q_L$ & $\nu_R$ & $e_R$ & $u_R$ & $d_R$ & $\xi$ \\
\hline 
$Y$ & $-\fr 1 2$ & $\fr 1 6$ & 0 & $-1$ & $\fr 2 3$ & $-\fr 1 3$ & $0$ \\ \hline
$N$ & $\delta-\fr 1 2 $ & $\fr 1 6-\fr{\delta}{3}$ & $\delta$ & $\delta -1$ & $\fr 2 3 -\fr{\delta}{3}$ & $-\fr 1 3-\fr{\delta}{3}$ & $2r$\\
\hline\hline
\end{tabular}
\caption[]{\label{alb} $Y,N$ quantum numbers of fermion multiplets in the generic case.}
\ec
\end{table}

All the anomalies are cancelled within each generation, independent of $\delta$, because of 
\bea
[SU(3)_C]^2 U(1)_Y &\sim & \sum_{\mathrm{quarks}} (Y_{f_L}-Y_{f_R}) = 3(2Y_q - Y_{u} - Y_{d})\crn
&=& 3[2(1/6)-(2/3)-(-1/3)]=0.\eea\bea
[SU(3)_C]^2 U(1)_N &\sim& \sum_{\mathrm{quarks}} (N_{f_L}-N_{f_R}) =  3(2N_q - N_{u} - N_{d})\crn
&=& 3[2(1/6-\delta/3)-(2/3-\delta/3)-(-1/3-\delta/3)]=0.
\eea
\be 
[SU(2)_L]^2 U(1)_Y \sim \sum_{\mathrm{doublets}} Y_{f_L}= Y_l+3Y_q = (-1/2)+3(1/6) =0.  
\ee 
\be
[SU(2)_L]^2 U(1)_N \sim \sum_{\mathrm{doublets}} N_{f_L}= N_l+3N_q = (-1/2+\delta)+3(1/6-\delta/3) =0.    
\ee
\bea 
[\mathrm{Gravity}]^2U(1)_Y&\sim&\sum_{\mathrm{fermions}}(Y_{f_L}-Y_{f_R})\crn
&=&2Y_l+2\times 3Y_q+Y_{\xi_L}-Y_{\nu}-Y_{e}-3Y_{u}-3Y_{d}-Y_{\xi_R}\crn
&=&2(-1/2)+6(1/6)+0-0-(-1)-3(2/3)-3(-1/3)-0=0.
\eea
\bea 
[\mathrm{Gravity}]^2U(1)_N&\sim&\sum_{\mathrm{fermions}}(N_{f_L}-N_{f_R})\crn
&=&2N_l+2\times 3N_q+N_{\xi_L}-N_{\nu}-N_{e}-3N_{u}-3N_{d}-N_{\xi_R}\crn
&=&2(-1/2+\delta)+6(1/6-\delta/3)+2r-\delta-(\delta-1)-3(2/3-\delta/3)\crn
&&-3(-1/3-\delta/3)-2r=0.
\eea
\bea 
[U(1)_Y]^2U(1)_N&=&\sum_{\mathrm{fermions}}(Y^2_{f_L}N_{f_L}-Y^2_{f_R}N_{f_R})=2Y^2_l N_l +2\times 3Y^2_q N_q +Y^2_{\xi_L}N_{\xi_L}\crn
&&-Y^2_{\nu}N_{\nu}-Y^2_{e}N_{e}-3Y^2_{u}N_{u}-3Y^2_{d}N_{d}-Y^2_{\xi_R}N_{\xi_R}\crn
&=& 2(-1/2)^2(-1/2+\delta)+6(1/6)^2(1/6-\delta/3)+0^2\times 2r-0^2\times \delta\crn
&& -(-1)^2(\delta-1)-3(2/3)^2(2/3-\delta/3)-3(-1/3)^2(-1/3-\delta/3)\crn
&& -0^2\times 2r=0.
\eea 
\bea 
U(1)_Y[U(1)_N]^2&=&\sum_{\mathrm{fermions}}(Y_{f_L}N^2_{f_L}-Y_{f_R}N^2_{f_R})=2Y_l N^2_l+2\times 3Y_q N^2_q+Y_{\xi_L}N^2_{\xi_L}\crn
&&-Y_{\nu}N^2_{\nu}-Y_{e}N^2_{e}-3Y_{u}N^2_{u}-3Y_{d}N^2_{d}-Y_{\xi_R}N^2_{\xi_R}\crn
&=& 2(-1/2)(-1/2+\delta)^2+6(1/6)(1/6-\delta/3)^2+0\times (2r)^2-0\times \delta^2\crn
&& -(-1)(\delta-1)^2-3(2/3)(2/3-\delta/3)^2-3(-1/3)(-1/3-\delta/3)^2\crn
&& -0\times (2r)^2=0.
\eea 
\bea 
[U(1)_Y]^3&=&\sum_{\mathrm{fermions}}(Y^3_{f_L}-Y^3_{f_R})\crn
&=& 2Y^3_l+2\times 3Y^3_q+Y^3_{\xi_L}-Y^3_{\nu}-Y^3_{e}-3Y^3_{u}-3Y^3_{d}-Y^3_{\xi_R}\crn
&=& 2(-1/2)^3+6(1/6)^3+0^3-0^3-(-1)^3\crn
&& -3(2/3)^3-3(-1/3)^3-0^3=0.\eea
\bea 
[U(1)_N]^3&=&\sum_{\mathrm{fermions}}(N^3_{f_L}-N^3_{f_R})\crn
&=& 2N^3_l+2\times 3N^3_q+N^3_{\xi_L}-N^3_{\nu}-N^3_{e}-3N^3_{u}-3N^3_{d}-N^3_{\xi_R}\crn
&=& 2(-1/2+\delta)^3+6(1/6-\delta/3)^3+(2r)^3-\delta^3-(\delta-1)^3\crn
&& -3(2/3-\delta/3)^3-3(-1/3-\delta/3)^3-(2r)^3=0.
\eea
Notice that the dark fermion $\xi$ is vector-like, not contributing to any anomaly, which need not necessarily be counted from the outset.

Additionally, as mentioned in the body text, if the model contains a variety of dark charges, say \be SU(3)_C\otimes SU(2)_L\otimes U(1)_Y\otimes U(1)_{N_1}\otimes U(1)_{N_2}\otimes \cdots \otimes U(1)_{N_p},\ee the anomalies of all types as computed above are still canceled. For the remaining anomalies, it is sufficiently to verify
\bea 
[U(1)_N]^2U(1)_{N'}&=&\sum_{\mathrm{fermions}}(N^2_{f_L}N'_{f_L}-N^2_{f_R}N'_{f_R})=2N^2_l N'_l +2\times 3N^2_q N'_q -N^2_{\nu}N'_{\nu}\crn
&&-N^2_{e}N'_{e}-3N^2_{u}N'_{u}-3N^2_{d}N'_{d}\crn
&=& 2(\delta-1/2)^2(\delta'-1/2)+6(1/6-\delta/3)^2(1/6-\delta'/3)-\delta^2\times \delta'\crn
&& -(\delta-1)^2(\delta'-1)-3(2/3-\delta/3)^2(2/3-\delta'/3)\crn
&&-3(-1/3-\delta/3)^2(-1/3-\delta'/3)=0,
\eea where the distinct values $\delta,\delta'$ define the hyperdark charges $N,N'$, respectively. Hence, the model of multi dark charges is viable, attracting attention.

\bibliographystyle{JHEP}
\bibliography{combine}
\end{document}